\documentclass[12pt]{article}
\usepackage{geometry}
\usepackage{a4}
\usepackage{graphics}
\usepackage{graphicx,subfigure}
\usepackage{subfigure}
\usepackage{epsf}
\usepackage[T1]{fontenc}
\usepackage[utf8]{inputenc}
\usepackage{amsmath}
\usepackage{amssymb}
\usepackage{cite}
\usepackage{tensor}

\newcommand{\be}{\begin{equation}}
\newcommand{\ee}{\end{equation}}
\newcommand{\bea}{\begin{eqnarray}}
\newcommand{\eea}{\end{eqnarray}}
\newcommand{\beal}{\begin{align}}
\newcommand{\eeal}{\end{align}}
\newcommand{\nn}{\nonumber}

\begin{document}
%%%%%%%%%%%%%%%%%%%%%%%%%%%%
\begin{titlepage}
\title{On subregion holographic complexity and renormalization group flows}
\author{}
\date{% authors are dated
Pratim Roy, Tapobrata Sarkar
\thanks{\noindent E-mail:~ proy, tapo@iitk.ac.in}
\vskip0.4cm
{\sl Department of Physics, \\
Indian Institute of Technology,\\
Kanpur 208016, \\
India}}
\maketitle\abstract{
\noindent
We investigate subregion holographic complexity in the context of renormalization group flow geometries. 
We use both the Poinca\'re slicing and the Janus ansatz as holographic duals to renormalization group flows in the boundary conformal field theory. 
In the former metric, subregion complexity is computed for a disc and a strip shaped entangling region. 
For the disc shaped region, consistent emergence of length scales for flow to the deep infra-red
is established. For strip shaped regions, we find that
complexity cannot locate holographic phase transitions in a sharp domain wall scenario. For smooth domain walls, we find that the
complexity might be an indicator of such phase transitions, and give numerical evidence that its derivative changes sign across a
transition. Finally, the complexity is computed numerically using the Janus ansatz.}
\end{titlepage}

\section{Introduction}

The statistics of information, popularly known as ``information theory'' is an old topic that has attracted great interest in diverse areas 
of science over the last few decades. Its more recent {\it avatar}, the quantum theory of information promises to be a leading character 
in the study of the quantum computer. While in its initial form, information theory was devised to deal with questions related to the ``closeness'' 
of statistical distributions, in quantum theories, it broadly attempts to understand the following question : ``how close are two quantum states.''  
For pure states, this question was addressed in \cite{provost1980riemannian}. This was following an earlier work, which had dealt with a similar 
issue in the context of thermal states in an equilibrium thermodynamic system (for a review, see \cite{ruppeiner1995riemannian}). The thermal
metric is known to be related \cite{Brody} to the Fisher information metric, popularly studied in the statistics community. 
Broadly, \cite{provost1980riemannian} developed a notion of a 
quadratic form, or a Riemannian metric on the space of parameters of the system where two pure quantum 
states are infinitesimally separated. 

In the context of many body quantum systems at zero temperature, such discussions are relatively recent. Indeed, as is known by now, 
there are several measures of information in this context, one of them being the proposal of \cite{zanardi2006ground}, that given two quantum 
ground states in a many-body system denoted by $|g_1\rangle$ and $|g_2\rangle$, the overlap function $\langle g_1|g_2\rangle$ provides a 
characterization of quantum phase transitions. In \cite{zanardi2007information}, the work of \cite{zanardi2006ground} was cast in a language
similar to that of \cite{provost1980riemannian} and a ``quantum information metric'' was constructed for the transverse XY spin chain in a 
homogeneous magnetic field. It was shown that this metric (or more appropriately the scalar curvature associated to it) was indeed an effective 
measure to characterize quantum phase transitions in the model. A flurry of activity followed soon after, and the information metric and a 
related quantity known as the fidelity susceptibility were calculated in a variety of examples and remain two of the most popular information theoretic 
quantities (for a review, see \cite{gu2010fidelity}). 

In quantum field theories, information theoretic studies have a long history, and it was realised sometime back that 
the renormalization group equations can be written in a geometric framework, see, e.g \cite{Lassig:1989tc,Dolan:1993cf,Dolan:1995zq,Dolan:1997cx}.
A natural issue that then arises is the application of similar ideas in the context of string theory. While a study of overlaps between string 
states (in lines with \cite{zanardi2006ground}) might require a deeper understanding of string field theory and could be somewhat complicated, 
nonetheless the anti-de-Sitter/conformal field theory (AdS/CFT) correspondence provides a new approach to understanding the origins of 
quantum information in the context of quantum field theories, via gravity duals. 

Indeed, there has been a recent upsurge of interest in using the AdS/CFT correspondence to elucidate several key concepts which are essentially 
quantum information theoretic in nature. Starting from the celebrated Ryu-Takayanagi (RT) conjecture for the entanglement entropy in quantum 
field theories \cite{TakayanagiMain}, one such issue of current interest is the complexity of a quantum state. This quantity may roughly be 
defined as the least possible number of steps needed to construct the state from a given reference (thus capturing the ``difficulty'' in creating 
a quantum state). There are currently several proposals in the literature for calculating quantum complexity using the dual holographic route. 

The first proposal, popularly known in the literature as the complexity = volume (CV) conjecture \cite{Susskind:2014rva, Susskind:2016tae} 
states that the complexity of a given quantum state on a time slice of the boundary CFT is dual to the volume of the maximal hypersurface in the 
bulk of co-dimension one, which is matched to the boundary at the given time. The second proposal is the complexity = action (CA) conjecture \cite{Brown}. 
This states that the holographic complexity can be evaluated by calculating the gravitational action on a Wheeler-de Witt patch of the bulk spacetime.
A related proposal, put forward in \cite{Alishahiha}, deals with holographic complexity (in the sense of a ``reduced fidelity susceptibility'')
of subregions of the boundary CFT (for related works, see e.g \cite{Benami,Momeni3,Wang,Gan1,Flory1,Momeni5}). This is in some sense a generalization of the
CV conjecture to specific subregions in the boundary and relates to the volume of the bulk spacetime enclosed by a Ryu-Takayanagi minimal surface 
(called RT surface in sequel) of co-dimension two. We will interchangeably call this the ``complexity'' or equivalently the  ``RT volume'' in sequel. 
In a more recent work, \cite{Souvik} attempts to relate the reduced fidelity susceptibility to the Fisher information metric, whose holographic
dual was calculated in \cite{Lashkari}.

In this paper, we consider renormalization group (RG) flow scenarios in the context of subregion holographic complexity. For the CFT, 
adding a relevant operator triggers an RG flow to either an IR conformal fixed point, or the theory becomes massive. We study holographic 
RG flow geometries using both the Poinca\'re slicing and the Janus ansatz. Similar studies have been undertaken for the entanglement 
entropy in the past (see, for example, \cite{Johnson,Mezei,Gutperle}). It is also relevant to mention that much of the 
work done in the context of the holographic entanglement entropy has been motivated by providing a holographic description for the 
c-functions and its proposed variants \cite{deBoer, MyersRG}. The Janus ansatz, which is the holographic dual of interface and 
boundary CFTS (ICFTS and BCFTs) has been treated in \cite{Gutperle}. Although complexity does not naturally have an analogue of 
c-functions, this study is physically interesting in its own right. One can ask if, for example, what physical features of the system are 
captured by subregion complexity along RG flows. In particular, a natural question is whether complexity
encodes information regarding holographic phase transitions (as discussed in \cite{MyersRG}). It is also interesting to understand the emergence of
different length scales in complexity along an RG flow, akin to what had been discussed in the context of entanglement entropy in \cite{Johnson}. 
Further, keeping in mind the proposal of \cite{Souvik}, this could give us a concrete realization of the Fisher information metric in 
multi-parameter systems.

With the above motivations in mind, we initiate a study of how holographic complexity behaves in a RG flow scenario. 
The paper is organized as follows: after a brief review in section 2 to fix the notations and conventions used in the paper, 
in section 3, we first compute the complexity for $d=3,4$ dimensional CFTs for a 
disc shaped entangling region (and assuming a sharp domain wall), for which the expressions are analytically simple and amenable to 
physical interpretation. We point out some differences between the complexity and the entanglement entropy in such a scenario and how different
scales emerge in the context of the RG flows. We then proceed to calculate the complexity for an abrupt domain wall type of geometry
for a strip shaped boundary subsystem and point out some of its aspects. It is known that this set-up exhibits 
a phase transition for certain values of the parameters and we track the behaviour of the complexity at this point.
Next, we generalize to a smooth domain wall and numerically calculate the complexity and analyze the issue of phase transitions in the system. 
Interestingly, we find numerical evidence that a derivative of the subregion complexity changes sign at the phase transition. 
In section 4, we consider the Janus ansatz, and compute the complexity of ICFT and BCFT numerically. 
Finally, in section 5, we conclude with a summary of results and future directions.

\section{Review and notation}

In this section, we will give a brief overview of the existing results in the literature on information theoretic geometry in the backdrop of AdS/CFT. 
This would serve to set the notations and conventions to be followed in the rest of the paper. Since the purpose of this section is to review the relevant
literature, we will be brief and details can be found in the references contained herein. 

As mentioned in the introduction, the quantity of interest in information geometry is the distance between two given quantum states. Say we have
a pure state $|\psi(q)\rangle$ where $q$ is a parameter in the theory.\footnote{In general, we could have $q \equiv \{q_i\}$ to denote a possible set of parameters.} 
Then, if we perturb the system by the infinitesimal change $q \to q + \delta q $, the new wave function for the initial pure state would be 
$|\psi(q + \delta q)\rangle$. The fidelity susceptibility $G_{qq}$, a quantity that we will be interested in this paper is then defined for the pure states as
\begin{equation}
|\langle \psi(q)|\psi(q+\delta q)\rangle| = 1 - G_{qq}(\delta q)^2 + \cdots
\label{FS}
\end{equation}
$G_{qq}$ is also popularly called as the Bures metric, and is distinct from the Fisher information metric between quantum states. 
In \cite{MIyaji:2015mia}, a gravity dual of the fidelity susceptibility was proposed for pure states of a $d+1$ 
dimensional boundary CFT in the context of the AdS/CFT correspondence. The proposal reads 
\begin{equation}
G_{qq} = \frac{{\rm Vol}\left(\Sigma_{max}\right)}{R^{d+1}}
\label{FSpure}
\end{equation}
up to an ${\mathcal O}(1)$ constant, where $\Sigma_{max}$ is a $d+1$ dimensional space-like bulk hypersurface that ends on the time slice at
the boundary of the AdS space and has maximal volume. In \cite{MIyaji:2015mia}, this was checked by explicitly computing the overlap between
two boundary states separated by a marginal deformation of the CFT on the field theory side. The analogue of eq.(\ref{FSpure}) for
mixed states is called quantum fidelity, and is defined for two density matrices $\rho_0$ and $\rho_1$ as 
\begin{equation}
F = {\rm Tr}\sqrt{\sqrt{\rho_0}\rho_1\sqrt{\rho_0}}
\label{FSmixed}
\end{equation}
which, for two separated pure states, reduce to their inner product. The fidelity susceptibility is the second derivative of the quantum fidelity
defined in eq.(\ref{FSmixed}). For reduced density matrices, this is more appropriately the reduced fidelity susceptibility. 

On the other hand, as stated in the introduction, the idea of a Riemannian metric on a quantum parameter manifold was first proposed in 
\cite{provost1980riemannian}. Given two pure states that are infinitesimally separated in a parameter space, this metric is given by a gauge
invariant combination and schematically reads 
\begin{equation}
g_{ij} = \langle\partial_i\psi|\partial_j\psi\rangle - \langle\psi|\partial_i\psi\rangle\langle\psi|\partial_j\psi\rangle
\label{qmetric}
\end{equation}
where the subscripts are derivatives with respect to the parameters of the theory $\{q_i\}$. This metric has important applications in 
the context of some quasi-fermionic systems, where the Hilbert space splits into a direct product of single particle states and hence has
been extensively studied for many body (pure) ground states. 
A natural extension of this metric in the realm of quantum field theory in the context of RG flows was first provided in \cite{Lassig:1989tc} and 
further extended by several authors. This metric was essentially derivatives of the free energy with respect to system parameters. 
For example, in \cite{Dolan:1995zq}, such metrics were defined for free field theories with a source (with the mass squared and the source coupling
providing the parameters), and have been since generalized to various other cases. We note here that choosing the correct coordinates on the 
parameter space is important. Often linear combinations (or otherwise) of system parameters provide simple insights into the 
nature of the metric on the parameter space. Scalar quantities such as the Ricci scalar might provide universal characterization, but 
might be difficult to handle in numerical analysis of the type that we will undertake in the later part of this paper. 

In the AdS/CFT context, the holographic dual of the Fisher information metric was first proposed in \cite{Lashkari}. For a ball shaped region
in a boundary CFT, defining 
\begin{equation}
\Delta S_B = S(\rho_B(q)) - S(\rho_B(0))~,~~ \Delta E_B = {\rm Tr}(H_B\rho_B(q)) - {\rm Tr}(H_B\rho_B(0))
\end{equation}
where $\rho$ denotes a density matrix, $S_B$ is the holographic entanglement entropy, $H_B$ is the modular Hamiltonian, and $q$ 
denotes a set of system parameters, it was shown that up to order $q$, one has
\begin{equation}
\frac{\partial^2}{\partial q^2}\left(\Delta E_B - \Delta S_B\right) = {\mathcal E} + \cdots 
\label{Fisherinfo}
\end{equation}
with ${\mathcal E}$ being the canonical energy \cite{Hollands} in classical gravity for perturbations in the Rindler wedge associated with 
the ball shaped region on a spatial slice of the boundary (with additional terms related to the Einstein tensor appearing at higher orders). 
The left hand side of the above equation is the Fisher information metric, a second derivative of the ``relative entropy'' that is a measure 
of the distinguishability of a density matrix with respect to another reference density matrix. 

Based on the CV conjecture of  \cite{Susskind:2014rva}, in \cite{Alishahiha} a proposal for the holographic complexity was given for specific
subregions of the boundary theory. This reads 
\begin{equation}
{\mathcal C}_{subregion} = \frac{V(\gamma)}{8\pi G L_{AdS}}
\end{equation}
where $V(\gamma)$ is the volume of a minimal Ryu-Takayanagi surface, $G$ 
is the Newton's constant in appropriate dimension, and $L_{AdS}$ is an AdS length scale. Various aspects of this quantity
has been studied in the literature, including low and high temperature expansions in a thermal setting (see, e.g \cite{tapo1}). 

In \cite{Souvik}, it was shown that for a certain class of examples, one can relate the Fisher information metric defined in eq.(\ref{Fisherinfo})
to a perturbatively regulated volume of the RT surface that corresponds to a given mixed state on the boundary, as the second derivative 
(with respect to the perturbing parameter) of the regulated RT volume. 

With this brief introduction, we now proceed to study holographic complexity in some RG flow scenarios. 

\section{Holographic RG flow geometries}
\label{RGflow}
As mentioned in the introduction, in the context of the AdS/CFT correspondence, an RG flow is triggered by a relevant operator in the CFT. 
The holographic dual to this operator is a bulk geometry which incorporates a scalar field. The resulting bulk action is given by (we follow the
notations of \cite{MyersRG}),
\be
\label{action}
S=\frac{1}{2\kappa^2} \int d^{(d+1)}x \sqrt{-g} (R-\frac{1}{2}(\partial \phi)^2-V(\phi))
\ee
It is assumed that the scalar potential $V(\phi)$ has stationary points such that $\left.\frac{\delta V}{\delta \phi} \right|_{\phi_i} =0$ where the 
spacetime is AdS, with a negative potential energy $V(\phi_i)=-\frac{d (d-1)}{L^2}\alpha_{i}^2$. Here, $L$ is the fundamental length scale 
in the theory and $\alpha_{i}$ are dimensionless constants which are different for various fixed points. At the fixed points, the AdS curvature 
scale is given by $\tilde{L}=L/\alpha_{i}$. It is well known that the metric representing a holographic RG flow is given by \cite{Freedman},
\be
\label{metricPoincare}
ds^2 = dr^2 + e^{2A(r)}\eta_{ij}dx^i dx^j
\ee
In these coordinates, the UV is located at $r \rightarrow \infty$ and the IR is at $r \rightarrow -\infty$. At both the IR and the UV fixed points, 
the conformal factor $A(r)$ takes the simple form $\frac{r}{\tilde{L}}$. Note that the AdS length $\tilde{L}$ is different for the IR and the UV, 
with $L_{UV} > L_{IR}$. Of course, it is necessary to introduce a UV cut-off $r_{\infty}$ to obtain a finite result in holographic calculations. 
This cut-off has the expression $r_{\infty}=L_{UV}\log(\frac{L_{UV}}{\delta})$, which is related to the standard cut-off $z_{min}=\delta$ for the AdS 
metric in the Fefferman-Graham coordinates \cite{MyersRG}.

We note here that  that the conditions $A(r) \sim r$ as $r \to \pm\infty$ and $A''(r) \leq 0$ (required to rule out further AdS 
boundaries \cite{Freedman}) are quite restrictive and one is limited to few consistent choices for $A(r)$. One of this is the step 
profile that we will consider in the next subsection, and an example of a continuous profile will be subsequently considered.
We will now calculate the volume $V_{\Sigma}$ of the RT surface for various forms of the factor $A(r)$ and for various types of entangling surfaces.
We begin with a disc shaped entangling region in a sharp domain wall geometry. Note that the sharp domain wall geometry is not
an exact supergravity solution. Nonetheless, this is a good model to study as it offers analytical handle which is difficult elsewhere. 

\subsection{A sharp domain wall: Disc shaped entangling region}

The explicit form for the conformal factor $A(r)$ is taken as \cite{Johnson},\cite{MyersRG},
 \be
 \label{stepprofile}
A(r)=\bigg\{
 \begin{array}{l l l}
A_{IR}(r)=\frac{r\,-\,r_0}{L_{IR}}+\frac{r_0}{L_{UV}} & \textrm{for }
r\leq r_0 &  \\
A_{UV}(r)=\frac{r}{L_{UV}} &\textrm{for }r\geq r_0 & \,,
\end{array}
 \ee
This is a step profile, which is continuous at $r=r_0$, the radius at which two AdS metrics with different curvatures are sewn together. 
We investigate the disc shaped entangling regions for the bulk metric of 
eq.(\ref{metricPoincare}) starting with (a ball shaped region in) $AdS_{5}$. We denote\footnote{$\rho$ is a boundary radial
coordinate here, and is not to be confused with the notation of the density matrix in the last section.}  by $r = r(\rho)$
the embedding of the ball shaped region in $AdS_5$, and take the entangling region 
to be $\rho \leq \ell$. Then the area of the RT surface that extends into the bulk in this case is given by,
\be
\text{Area}=4\pi \int_{0}^{\ell} d\rho \rho^2 e^{3A(r)} \bigg(1+e^{-2A(r)}r^{\prime}(\rho)^2 \bigg)^{\frac{1}{2}} 
\ee
The solution to the equations of motion are obtained by minimizing this area functional. Introducing an UV cut-off $\delta$, which cuts off
the integral for $r$ at $r_{\infty} = - L_{UV}\log(\delta/L_{UV})$, this solution is given by \cite{Johnson}
 \be
 \label{eomsol}
r(\rho)=\bigg\{
 \begin{array}{l l l}
r_{IR}(\rho)=-\frac{L_{IR}}{2}\log\bigg(\frac{\ell^2+\delta^2-\rho^2+c_{IR}^2}{L_{IR}^2} \bigg) & \textrm{for }
\rho \leq \rho_{0} &  \\
r_{UV}(\rho)=-\frac{L_{UV}}{2}\log\bigg(\frac{\ell^2+\delta^2-\rho^2}{L_{UV}^2} \bigg) &\textrm{for }\rho > \rho_{0} & \,,
\end{array}
 \ee
 where $r(\rho_{0})=r_{0}$ and position of the domain wall is given by,
 \be
 \rho_{0}=\sqrt{\ell^2+\delta^2-L_{UV}^2 e^{-2 r_{0}/L_{UV}}}
 \ee
The constant $c_{IR}=\sqrt{L_{IR}^2e^{-2 r_{0}/L_{IR}}-L_{UV}^2 e^{-2 r_{0}/L_{UV}}}$ is determined by the 
requirement $r_{UV}(\rho_{0})=r_{IR}(\rho_{0})$.
We find that as in the case of the entanglement entropy in this setup, the RT volume comes out in terms 
of some combinations of parameters, and 
for convenience, we define an effective length scale
\be
\ell_{eff} = \sqrt{\ell^2 - L_{UV}^2 e^{-2 r_{0}/L_{UV}}} \equiv \sqrt{\ell^2 - \ell_{cr}^2+ {\mathcal O}(\delta^2)}
\label{leff}
\ee
which is reminiscent of an effective radius of the entangling disc as seen from the IR, with $\ell_{cr}^2 = L_{UV}^2 e^{- 2r_{0}/L_{UV}} + 
{\mathcal O}(\delta^2)$ being the minimum disc radius for the minimal surfaces to penetrate into the IR \cite{Johnson}.
The equation which gives the expression for the volume is given schematically as,
\begin{equation}
V_{\Sigma} = 4\pi \int_{0}^{\ell} d\rho \rho^2 \int_{r(\rho)}^{r_\infty} dr e^{3A(r)}  \\
\end{equation}	
Recall, at this point that the definition of the conformal factor (eq.(\ref{stepprofile})) and the subsequent solutions for the 
profile of the RT surface given by $r(\rho)$ (see eq.(\ref{eomsol})). The volume $V_{\Sigma}$ is the sum of contributions 
from both the UV and IR regions, which are evaluated as separate terms,
\begin{eqnarray}
V_{\Sigma} &=& {4\pi \int_{0}^{\rho_{0}}d\rho \rho^2\left[
\int_{r_{IR}(\rho)}^{r_{0}} dr e^{3A_{IR}(r)}+ \int_{r_0}^{r_{\infty}}dre^{3A_{UV}(r)}\right]}\nonumber\\
&+& 4\pi \int_{\rho_{0}}^{\ell} d\rho \rho^2 \int_{r_{UV}(\rho)}^{r_{\infty}} dr e^{3A_{UV}(r)}
\label{volnew}
\end{eqnarray}	
Now, the volume as computed from eq.(\ref{volnew})
%\bea
%V_{\Sigma} &=& 4\pi \int_{0}^{\ell} d\rho \rho^2 \int_{r(\rho)}^{r_{\infty}} dr e^{3A(r)}  \\ \nn
% &=& 4\pi \int_{0}^{\rho_{0}}d\rho \rho^2 \int_{r_{IR}(\rho)}^{r_{0}} dr e^{3A_{IR}(r)} + 4\pi \int_{\rho_{0}}^{\ell} d\rho 
 %\rho^2 \int_{r_{UV}(\rho)}^{r_{\infty}} dr e^{3A_{UV}(r)}  
%\eea 
evaluates to a lengthy expression, which is best presented by defining\footnote{As explained in 
\cite{Johnson}, ${\tilde\epsilon}$ plays a similar role as the UV cutoff $\delta$, however, it does not necessarily have to be small.}
${\tilde\epsilon} = L_{IR}e^{-r_0/L_{IR}}$, 
in which case it reads, in terms of $\ell_{eff}$ of eq.(\ref{leff}) :
\bea
\label{discAdS5}
V_{\Sigma} &=& \frac{2\pi^2 L_{UV}^4}{3}+\frac{4\pi L_{UV}^4}{9}\left(\frac{\ell}{\delta}\right)^3 
-\frac{4\pi L_{UV}^4}{3}\left(\frac{\ell}{\delta}\right)\nonumber\\
&-&\frac{4 \pi L_{UV}^4}{9}\left(\frac{\ell_{eff}}{\ell_{cr}}\right)^3+ \frac{4\pi}{9}L_{IR}^4 \left(\frac{\ell_{eff}}{{\tilde\epsilon}}\right)^3 
+\frac{4\pi}{3}L_{UV}^4\left(\frac{\ell_{eff}}{\ell_{cr}}\right) - \frac{4\pi}{3} L_{IR}^4 \left(\frac{\ell_{eff}}{{\tilde\epsilon}}\right)\nonumber\\
&-&\frac{4\pi}{3}L_{UV}^4 \tan^{-1} \left(\frac{\ell_{eff}}{\ell_{cr}} \right) 
+\frac{4\pi}{3}L_{IR}^4 \tan^{-1} \left(\frac{\ell_{eff}}{{\tilde\epsilon}} \right)  + {\mathcal O}(\delta)
\eea
We also record here the volume of the RT surface in the case when the geodesics do not penetrate into the IR, i.e lie purely in the UV. In
this case, it is easy to show that this volume is simply the first line of eq.(\ref{discAdS5}), i.e
\bea
\label{AdS5discUV}
V_{\Sigma}^{UV} = \frac{2\pi^2 L_{UV}^4}{3}+\frac{4\pi L_{UV}^4 \ell^3}{9\delta^3}-\frac{4\pi L_{UV}^4 \ell}{3\delta}
\eea
Expectedly, this is also the result for $\ell_{eff} = 0$, i.e when the geodesics do not penetrate into the IR. 

There are a couple of things to be noted here. First, the divergence structure of the volume in the above expression is $\sim {\mathcal O}(\delta^{-3}) + 
{\mathcal O}(\delta^{-1})$, which is as expected for a bulk $AdS_5$, and gives the familiar 
``volume law'' (the strongest divergence structure). 
Secondly, let us consider the limit $\ell_{eff}/{\tilde\epsilon}\gg 1$, i.e the effective length is much greater than the IR cutoff. 
Then, it is not difficult to see that the terms involving $L_{IR}$ in eq.(\ref{discAdS5}) precisely go over
to their UV counterparts in the first line of that equation, with the replacement $(\ell/\delta) \to (\ell_{eff}/{\tilde\epsilon})$. 
This is what one would obtain if one is purely in the IR. 

Before we end this subsection, we briefly comment on the case of bulk $AdS_4$. A similar calculation for $AdS_4$ yields 
(with $\ell_{cr}$, $\ell_{eff}$, and $\tilde\epsilon$ being defined in the same way as in the $AdS_5$ case), 
\bea
\label{discAdS4}
V_\Sigma &=& \frac{\pi L_{UV}^3 }{2}\left(\frac{\ell}{\delta}\right)^2 -\pi L_{UV}^3 \log\left(\frac{\ell_{cr}}{\delta} \right) \nonumber\\
&-& \frac{\pi L_{UV}^3}{2} \left( \frac{\ell_{eff}}{\ell_{cr}} \right)^2
+ \frac{\pi L_{IR}^3}{2} \left(\frac{\ell_{eff}}{\tilde{\epsilon}} \right)^2
 -\frac{\pi}{2} L_{IR}^3 \log \left(1+\frac{\ell_{eff}^2}{{\tilde\epsilon}^2}\right)
\eea
and the RT volume for geodesics that lie purely in the UV region and does not see the domain wall is a purely divergent piece,
\bea
\label{AdS4discUV}
V_{\Sigma}^{UV}= \frac{\pi}{2} L_{UV}^3 \left(\frac{\ell}{\delta}\right)^2 -\pi L_{UV}^3 \log\left(\frac{\ell}{\delta} \right)
\eea
Let us note some salient features of these equations as well. The ``volume law'' is again satisfied, as expected.
Next, we note that for $\ell_{eff} = 0$, the pure UV contribution matches with the first line of eq.(\ref{discAdS4}). 
Then, we consider the case $\ell/\ell_{cr} \sim 1$. Again, if we assume in the spirit of our previous discussion
that $\ell_{eff}/\tilde\epsilon\gg 1$, then we see that the terms involving $L_{IR}$ are again similar in form to their UV cousins in
the first line of eq.(\ref{discAdS4}), with the replacement $(\ell/\delta) \to (\ell_{eff}/{\tilde\epsilon})$.

\subsection{A sharp domain wall : Strip shaped entangling Region}

We now briefly consider a sharp domain wall with a strip shaped entangling region. In this subsection, we will restrict ourselves to a bulk $AdS_3$, 
and take the entangling surface to be a strip of length $\ell$. This geometry was considered in \cite{MyersRG}, where it was shown that
this geometry shows an interesting phase behaviour with a holographic phase transition (as a function of the strip length).
In the initial part of this subsection,
we briefly review the results contained therein. We remind the reader that the conformal factor $A(r)$ is taken to be of the form of eq.(\ref{stepprofile}) 
with the location of the domain wall at $r=r_0$, and as mentioned in the previous subsections, this profile is not a solution of the Einstein's 
equations resulting from eq.(\ref{action}), but we use it here as a toy model in conjunction with a strip shaped entangling region. 
We record the expression for the area functional,
\be
A=\int_{0}^{(\ell-\epsilon)/2}dx \sqrt{(r^\prime)^2+e^{2A(r)}}
\ee
The fact that the integrand does not depend on $x(r)$ translates into 
\be
\frac{dx}{dr} = \frac{e^{-2A(r)}}{\sqrt{K_2^2-e^{-2A(r)}}}
\label{dxdrgeneral}
\ee
where we will consider only the branch for which the derivative of eq.(\ref{dxdrgeneral}) is positive. 
Here, $K_2 \equiv K_2(\ell)$ is a conserved quantity that is determined by demanding $x=0$ at $r=r_*$, which is the turning point of the 
minimal surface. There are two types of minimal surfaces for the geometry denoted by eqs. (\ref{metricPoincare}) and (\ref{stepprofile}), 
namely those surfaces that stay purely in the UV region and those that penetrate deeper into the bulk and reach the IR region. For the ones
that are entirely in the UV region, eq. (\ref{dxdrgeneral}) is integrated to give,
\be
\label{xpureUV}
x(r)=L_{UV}\sqrt{K_{UV}^2-e^{-2r/L_{UV}}}
\ee
According to the boundary condition mentioned before, $K_{UV} = e^{-r_*/L_{UV}}$. One also demands that $x=\ell/2$ as $r \rightarrow \infty$, 
so that $\ell = 2L_{UV}e^{-r_0/L_{UV}}$. Since the turning point of the minimal surface lies in the UV region, we may define a relation 
$\ell \leq \ell_2$, where $\ell_2 = 2L_{UV}e^{-r_0/L_{UV}}$. 

The second, and more interesting class of surfaces, penetrate into the IR, implying that $r_*<r_0$. The conserved quantity in this case is,
\be
K_2=K_{IR}=e^{-A_{IR}(r_*)}
\ee
The value of $x$ for which the minimal surface encounters the domain wall is denoted by, $x_t=x_{IR}(r_0)=x_{UV}(r_0)$. The parts of the 
geodesic lying in the UV and IR regions ($x_{UV}$ and $x_{IR}$ respectively) may readily be found by directly integrating 
eq.(\ref{dxdrgeneral}) to obtain 
\bea
x_{IR} = L_{IR}\sqrt{K_{IR}^2-e^{-2A_{IR}(r)}}~,~~ 
x_{UV} = L_{UV} \sqrt{K_{IR}^2-e^{-2A_{UV}(r)}}+c_{UV}
\eea
Here, $c_{UV}$ is an integration constant. By imposing the conditions, $x_{IR}(r_*)=0$ and $x_{UV}(r)=\ell/2$, $c_{UV}$ is found to be,
\be
c_{UV}=\frac{\ell}{2}-L_{UV}K_{IR}
\ee
Expressions for the length of the strip and location $x_t$ of the domain wall may be found from the above equations and give
\bea
\ell &=& 2L_{UV}K_{IR}-2(L_{UV}-L_{IR})\sqrt{K_{IR}^2-e^{-2r_{0}/L_{UV}}}\nonumber\\
x_t &=& \frac{L_{IR}}{L_{UV}-L_{IR}}\bigg(L_{UV}K_{IR}-\frac{\ell}{2} \bigg)
\label{lxtsdw}
\eea
From this, we may solve for $K_{IR}$ to obtain,
\be
\label{KIRroots}
K_{IR\pm} =\frac{L_{UV}\ell \pm \sqrt{\ell^2-4 L_{IR}(2L_{UV}-L_{IR})e^{-2r_0/L_{UV}}}}{2L_{IR}(L_{UV}-L_{IR})}
\ee
It may be noted that for the $K_{IR_{\pm}}$ to have real roots, we must have $\ell \geq \ell_{cr}$ with,
\be
\ell_{cr} \equiv 2\sqrt{L_{IR}(2L_{UV}-L_{IR})e^{-2r_0/L_{UV}}}
\label{lcr}
\ee
Here, $\ell_{cr}$ is interpreted as the critical strip length so that the geodesics penetrate into the IR. This is similar to the critical 
length encountered for the disc shaped entangling region in the previous subsections. 

From the above discussions, it follows that as far as the entanglement entropy is concerned, in the region $\ell_{cr} < \ell < \ell_2$, 
there are three valid solutions of the entanglement entropy. The pure UV part of the entropy is valid for $\ell < \ell_{2}$,  
whereas for $\ell > \ell_{2}$, the root $K_{IR+}$ correspond to consistent solutions, with $K_{IR-}$ being ruled out as
$dx/dr$ becomes negative here. 

With these necessary ingredients from \cite{MyersRG}, we will now set up the expression for the volume enclosed by minimal RT 
surface given from eq.(\ref{metricPoincare}). The volume is given by an expression analogous to eq.(\ref{volnew}), and reads
\bea
&~&V_{\Sigma} = \int dx dr e^{A(r)}  \nonumber\\
&=& \int_0^{x_t} dx \left[\int_{r_{IR(x)}}^{r_0} dr e^{A_{IR}(r)} + \int_{r_0}^{r_{\infty}} dr e^{A_{UV}(r)}\right] \nonumber\\
&+& \int_{x_t}^{(\ell-\epsilon)/2}dx \int_{r_{UV(x)}}^{r_{\infty}}dr e^{A_{UV}(r)}
\label{volsdw}
\eea
Here, the UV cut-off $r_{\infty} = L_{UV} \log(L_{UV}/\delta)$. Note that we are using a separate cut-off $\epsilon$ for $x$, which 
may simply be related to the UV cut-off $\delta$ by the relation $r(x=(\ell-\epsilon)/2)=r_{\infty}$ yielding \cite{MyersRG},
\be
\epsilon = \frac{\delta^2}{L_{UV}K_{IR}}
\ee
It is easier to present the results by setting, without loss of generality, $r_0=0$, and with this, the expression for the complexity reads
\begin{eqnarray}
&~&V_{\Sigma} = \frac{L_{UV}^2 \ell}{2\delta} + \frac{L_{IR}}{2} \left(\ell-2K_{IR} L_{UV} \right) \nonumber\\
&+& \left(L_{UV}^2 - L_{IR}^2 \right) \tan^{-1} \left( \frac{2K_{IR} L_{UV}-\ell}{\sqrt{(2K_{IR}L_{IR}-\ell)(2K_{IR}(L_{IR}-2L_{UV})+\ell}} \right)
\label{VolAdS3final}
\end{eqnarray}
Note that the divergent piece comes purely from the UV contribution (this can also be understood by evaluating the volume in 
the case that the geodesics do not penetrate into the IR). One can now use eq.(\ref{KIRroots}) to compute the complexity. 
Note that in the case of entanglement entropy, the pure UV contribution $\sim \log(\ell/\delta)$, so that one could consistently subtract
out the divergent piece by considering, for example, the difference in the entropy for two different strip lengths. In our case, it is more
useful to look at the finite part of the volume under the RT sufrace, by subracting the divergent part from eq.(\ref{VolAdS3final}). 

The behaviour of the complexity can be easily gleaned by substituting the values of $K_{IR\pm}$ from eq.(\ref{KIRroots})
in the finite part of eq.(\ref{VolAdS3final}). These are called  $V_{\Sigma+}$ and  $V_{\Sigma-}$ in sequel. 
From eq.(\ref{KIRroots}), it is seen that these match at $\ell = \ell_{cr}$ and that at $\ell = \ell_2$, $V_{\Sigma-}=0$. Also, it can
be checked that $V_{\Sigma+} - V_{\Sigma-}$ is a monotonically increasing function in the interval $\ell_{cr} < \ell < \ell_2$. 
From considerations of the entanglement entropy, it has been shown in \cite{MyersRG} that there is a first order phase transition
in the interval $\ell_{cr} < \ell < \ell_2$, when the entropy from the pure UV part equals that coming from the branch $K_{IR+}$. 
However, our discussion above coupled with the fact that in this case, there is no non-trivial dependence of the complexity on
the strip length in the pure UV region, leads to the conclusion that complexity does not capture information regarding this phase
transition, and that there is a discontinuous jump in the complexity at such a transition.

\subsection{A smooth domain wall: Strip shaped entangling region}

So far, we have just considered a sharp domain wall for simplicity. We may now consider a more realistic profile for the warp 
factor $A(r)$, which has also been considered in \cite{MyersRG}. Our aim in this subsection is to understand the nature of the RT volume
for these theories and contrast them with the entanglement entropy wherever possible. The importance of the smooth domain wall
profile in holographic RG flows is that this should remove the inconsistency that we have seen in the last two subsections, namely 
a position dependent nature of the divergence structure, that hampered a consistent renormalization of the complexity along the RG flow. 
We see in a moment that this is indeed true. 

\begin{figure}[h!]{
\begin{tabular}{cc}
\includegraphics[width=0.5\textwidth]{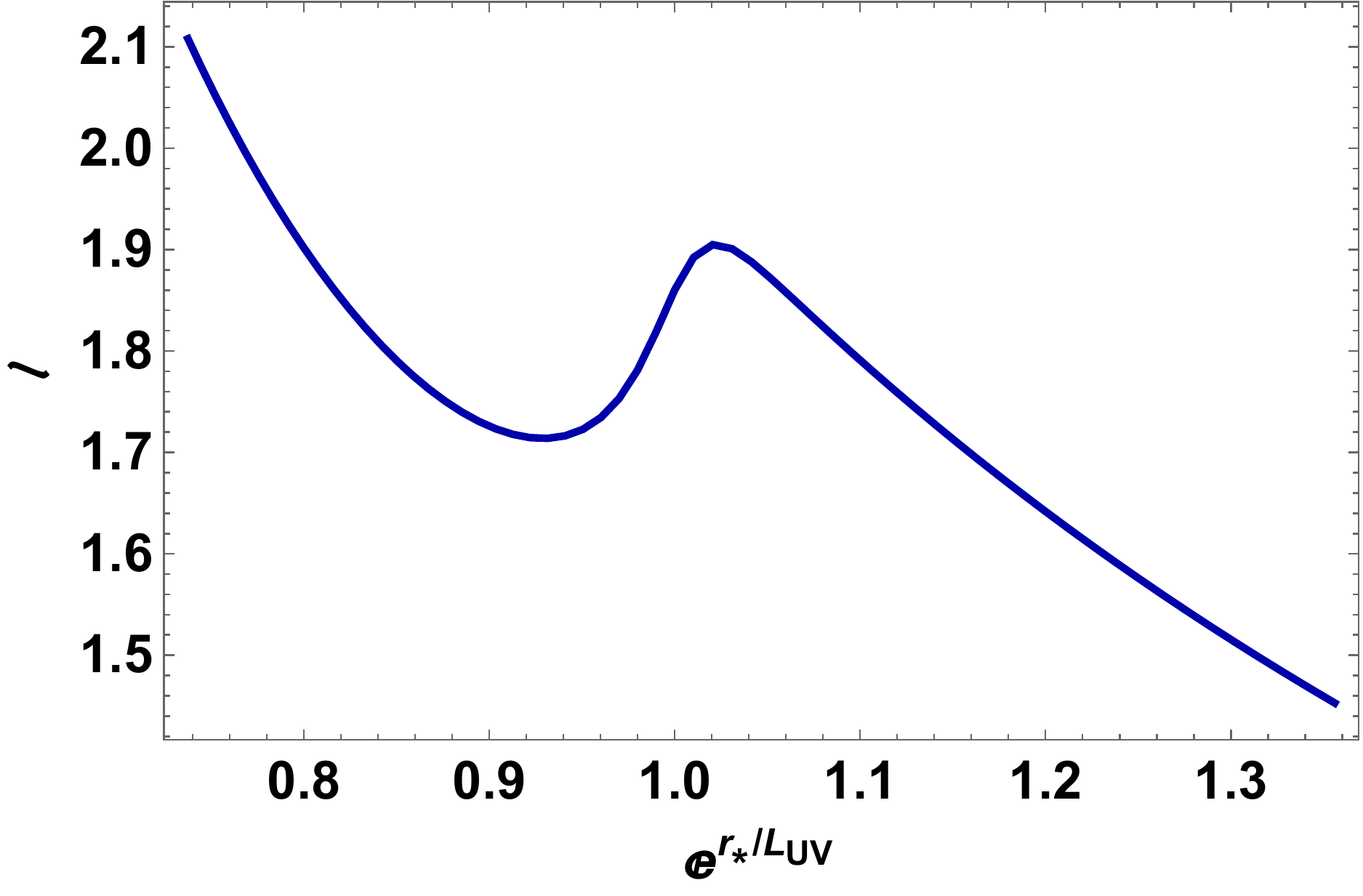}&
\includegraphics[width=0.5\textwidth]{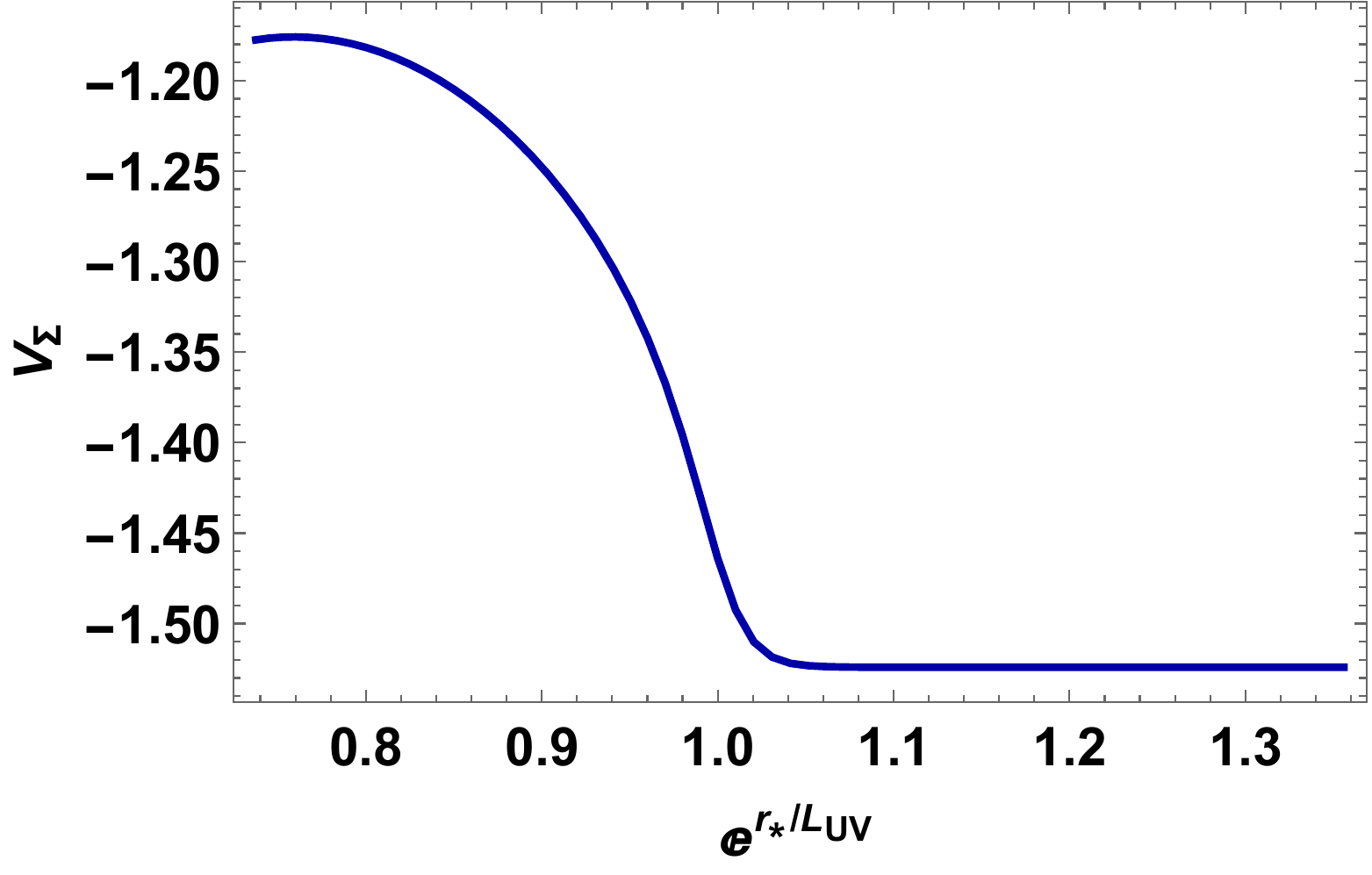}\\
(a) & (b)
\end{tabular}
\caption{Panel (a) $\ell$ is plotted as a function of $e^{r_{*}/L_{UV}}$. Panel (b) $V_{\Sigma}$ is plotted as a function of $e^{r_{*}/L_{UV}}$. 
Both (a) and (b) have the following parameter values: $L=0.66, \gamma=0.5R, R=0.02$. 
\label{llogVrstar}}}
\end{figure}

For a smooth domain wall, the functional form chosen for the profile is \footnote{the symbol $L$ used in eq.(\ref{smoothprofile})
should not be confused with $L$ or $L_{UV/IR}$ used previously in this section.}
\be
\label{smoothprofile}
e^{A(r)} = e^{r/L}(2 \cosh(r/R))^{-\gamma}
\ee
The AdS length scales at the UV and IR fixed points are given by,
\bea
\frac{1}{L_{UV}} = \left(\frac{1}{L} - \frac{\gamma}{R}\right) ~,~
\frac{1}{L_{IR}} = \left(\frac{1}{L} + \frac{\gamma}{R}\right)
\eea
As before, the equation determining the minimal surfaces is given by,
\be
\frac{dx}{dr} = \frac{e^{-2A(r)}}{\sqrt{K_{2}^2-e^{-2A(r)}}}
\ee
With the $A(r)$ chosen according to eq. (\ref{smoothprofile}), an analytic solution to the above equation is no longer possible. However, 
one determine that $x(r) \sim \sqrt{r-r_{*}}$ as $r \rightarrow r_{*}$. With this asymptotic behaviour, we may develop a series 
solution of $x(r)$ in the neighbourhood of the point $r=r_{*}$. The series solution can be treated as the initial condition for numerically finding 
the solution of $x(r)$ by shooting from a point close to $r=r_{*}$. For investigating the existence of a phase transition, it is necessary to plot the strip length 
$\ell$ as a function of $r_{*}$. We may obtain $\ell$ from the relation,
\be
\ell(r_{*}) = 2 \lim_{r \rightarrow \infty} x(r)
\ee
As discussed in \cite{MyersRG}, specific choices of the parameters $L$, $\gamma$ and $R$ can produce different types of solutions for the
strip length $\ell$. In particular, we will focus here on a choice of parameters that can produce multiple valued $\ell$ (for some range of
$r_{*}$) as a function of $r_{*}$. It is known that such solutions of $\ell$ might indicate first order phase transitions in the theory, as a function 
of the strip length. In fig. \ref{llogVrstar}(a), we present such a solution, for $L=0.66, \gamma=0.5R, R=0.02$ ($\gamma/R$ is 
fixed).\footnote{For ease of comparison with existing literature, we choose the same values of the parameters as in \cite{MyersRG}.} 
As we will see, the results are independent of the UV cut-off $\delta$. 
% % % % % % % % % % % % % % % % % % % % % % % % % % % % % % % % % % % % % % % % % % % % % % % % % 
% % % % % % % % % % % % % % % % % % % % % % % % % % % % % % % % % % % % % % % % % % % % % % % % % 
It may be seen from the fig.\ref{llogVrstar}(a) that the specific values of the parameters chosen produces a kink in $\ell$ for a range of $r_{*}$. 
The physical consequence of this kink is that there are multiple values of $r_{*}$ that are valid for a particular value of $\ell$ 
within a certain range of the strip length. This, of course is analogous to the range of $\ell$ in the previous subsection 
($\ell_{cr} \leq \ell \leq \ell_{2}$) for which there are multiple choices of the minimal surface. In the present case, we need to 
determine the values of $\ell_{cr}$ and $\ell_{2}$ numerically. 

The values of $\ell_{cr}$ and $\ell_{2}$ can be determined from the solution of $\ell (r_{*})$ by solving $\frac{d \ell}{d r_{*}}=0$ 
numerically. Having solved for $\ell_{cr}$ and $\ell_{2}$, we are in a position to compute the volume numerically from the integral,
\be
\label{smoothvolume}
V_{\Sigma} = \int_{r(x)}^{r_{\infty}} dr e^{A(r)}x(r)
\ee
The expression for the volume implicitly depends on $r_{*}$, since $\ell \equiv \ell(r_{*})$.
Now, since we have both $V_{\Sigma}$ and $\ell$ as a function of $e^{r_{*}/L_{UV}}$ (which is a convenient scale for the $x$-axis), 
it is not difficult to express $V_{\Sigma}$ as a function of $\ell$. A word on regulating the volume is in order. It can be verified that the factor 
$A(r)$ behaves as $A(r)\sim \frac{r}{L}-\frac{\gamma r}{R}$ as we approach $r \rightarrow \infty$. Evaluating the above integral in the 
limit $r \rightarrow \infty$ (which is responsible for the UV divergence in the volume) yields the divergent contribution,
\be
V_{\Sigma, div} = \frac{\ell}{2} \frac{1}{\left(\frac{1}{L}-\frac{\gamma}{R}\right)} e^{r_{\infty}\left(\frac{1}{L}-\frac{\gamma}{R}\right)}
\ee
that has to be subtracted out from eq. (\ref{smoothvolume}) to obtain a finite expression.
Indeed, this is the only divergence in the theory and has the expected $\ell/\delta$ behaviour. There is therefore no problem in
regulating the complexity (numerically) in a consistent manner throughout the flow. 
% % % % % % % % % % % % % % % % % % % % % % % % % % % % % % % % % % % % % % % % % % % % % % % % % 
\begin{figure}[h!]{
\begin{tabular}{cc}
\includegraphics[width=0.5\textwidth]{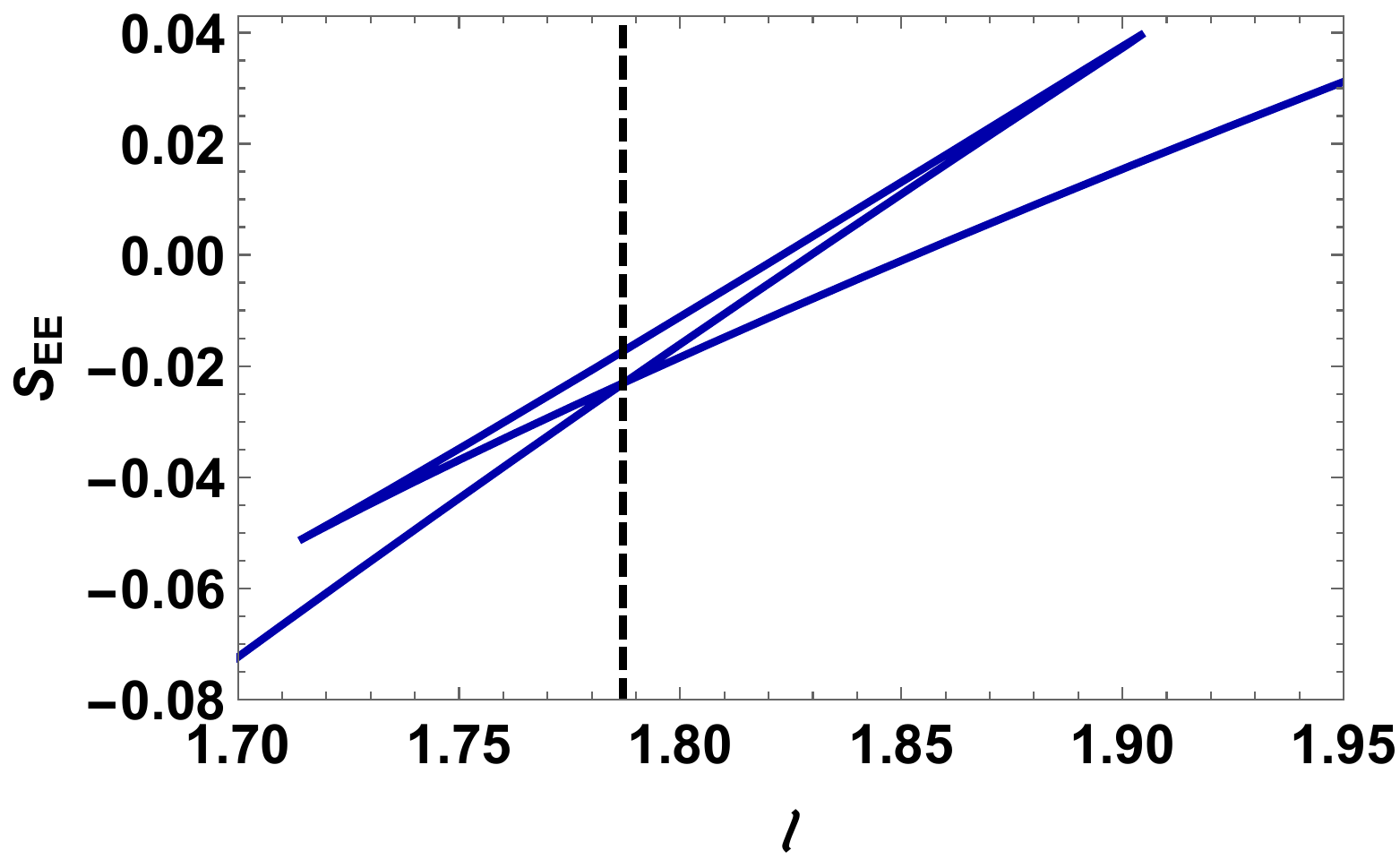}&
\includegraphics[width=0.5\textwidth]{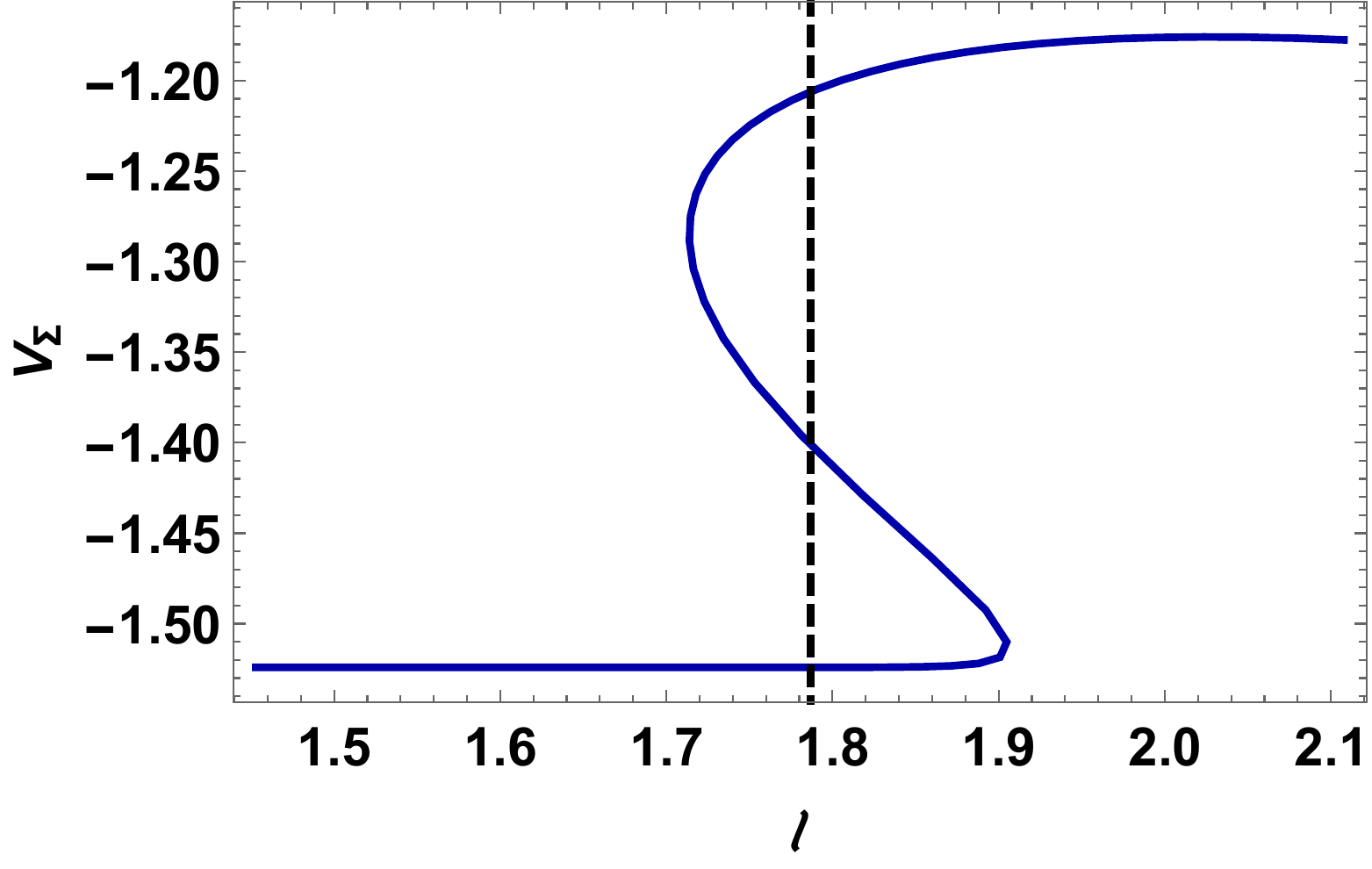}\\
(a) & (b)
\end{tabular}
\caption{Panel (a) Entanglement entropy is plotted as a function of $\ell$. Panel (b) $ V_{\Sigma}$ is plotted as a function of $\ell$. 
Both (a) and (b) have the following parameter values: $L=0.66, \gamma=0.5R, R=0.02$. The dashed
vertical line denotes the point of phase transition $\ell \approx 1.79$.} \label{SandV}}
\end{figure}
% % % % % % % % % % % % % % % % % % % % % % % % % % % % % % % % % % % % % % % % % % % % % % % % % 
As we have mentioned, it may be observed from fig. \ref{llogVrstar}(a) that there are three possible 
branches of $r_*$ for a fixed value of $\ell$ (for
a range of $\ell$). Fig. \ref{llogVrstar}(b) shows $V_{\Sigma}$ as a function of $e^{r_{*}/L_{UV}}$. Figs. \ref{SandV}(a) and (b)
show the entanglement entropy and the complexity as a function of $\ell$. In case of the former, we see the typical swallow-tail
behaviour as observed for the sharp domain wall in \cite{MyersRG}. As far as $V_{\Sigma}$ is concerned, it shows a multi-valued
behaviour with respect to $\ell$. In this figure, the vertical dashed line denotes the phase transition, $\ell \approx 1.79$.

It is more interesting to consider the variation of $V_{\Sigma}$ with $\gamma$ of eq.(\ref{smoothprofile}). This is a perturbation
parameter that appears in the potential (see eq.(C.8) of \cite{MyersRG}). In the spirit of our discussion in section 2, $\gamma$ can 
therefore be taken as a coordinate in the parameter space. The question that we specifically ask is whether there is any special behavior
of $V_{\Sigma}$ with respect to $\gamma$ close to the phase transition. This is interesting for the following reason. 

As discussed in the beginning of this paper, it is well known that information theoretic quantities like
the fidelity susceptibility or the Riemann curvature associated with the information metric typically show ``special'' behaviour 
near phase transitions (see, e.g \cite{zanardi2006ground}, \cite{zanardi2007information},\cite{gu2010fidelity}).\footnote{
For example, it is known that in the context of the one dimensional XY spin chain \cite{zanardi2007information}, 
the Riemann curvature computed from the information metric diverges at an anisotropic transition and suffers a discontinuity 
at the Ising transition.} It is therefore not difficult to envisage that the holographic subregion complexity that we computed 
might show some interesting behaviour at the phase transition discussed above. We give evidence below that the derivative of 
the complexity changes sign across this transition. 

\begin{figure}[h!]{
\begin{tabular}{cc}
\includegraphics[width=0.5\textwidth]{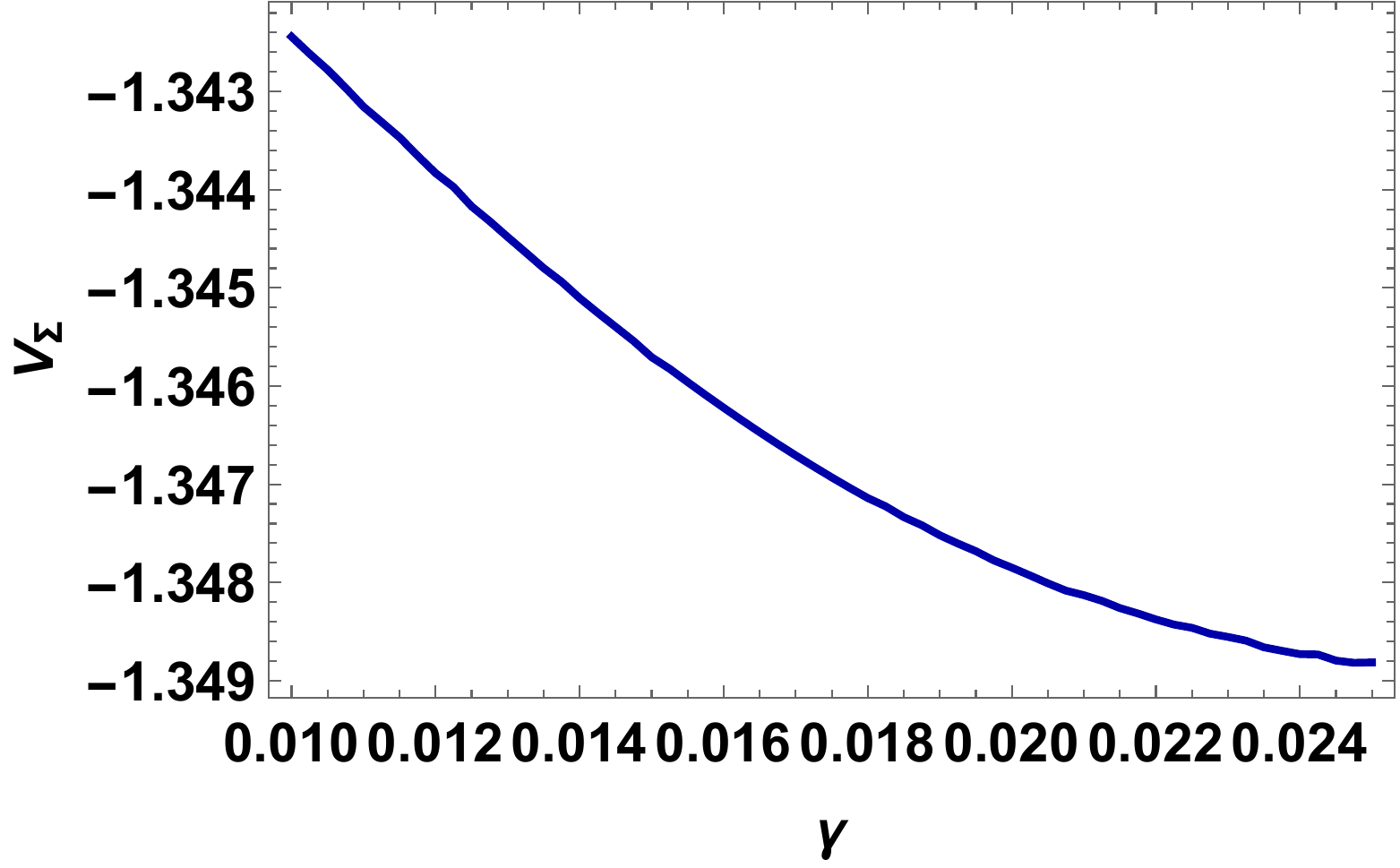}&
\includegraphics[width=0.5\textwidth]{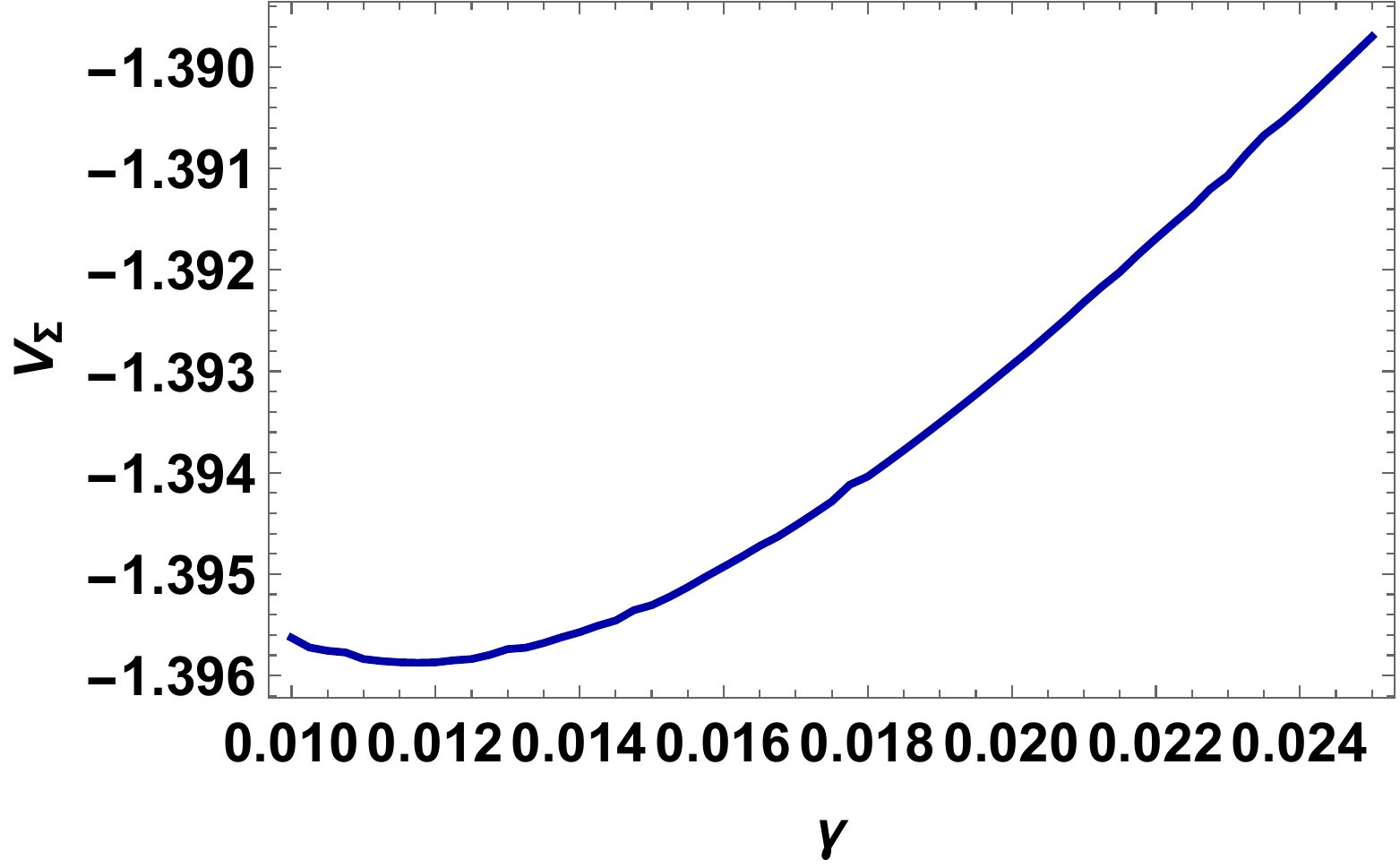}\\
\includegraphics[width=0.5\textwidth]{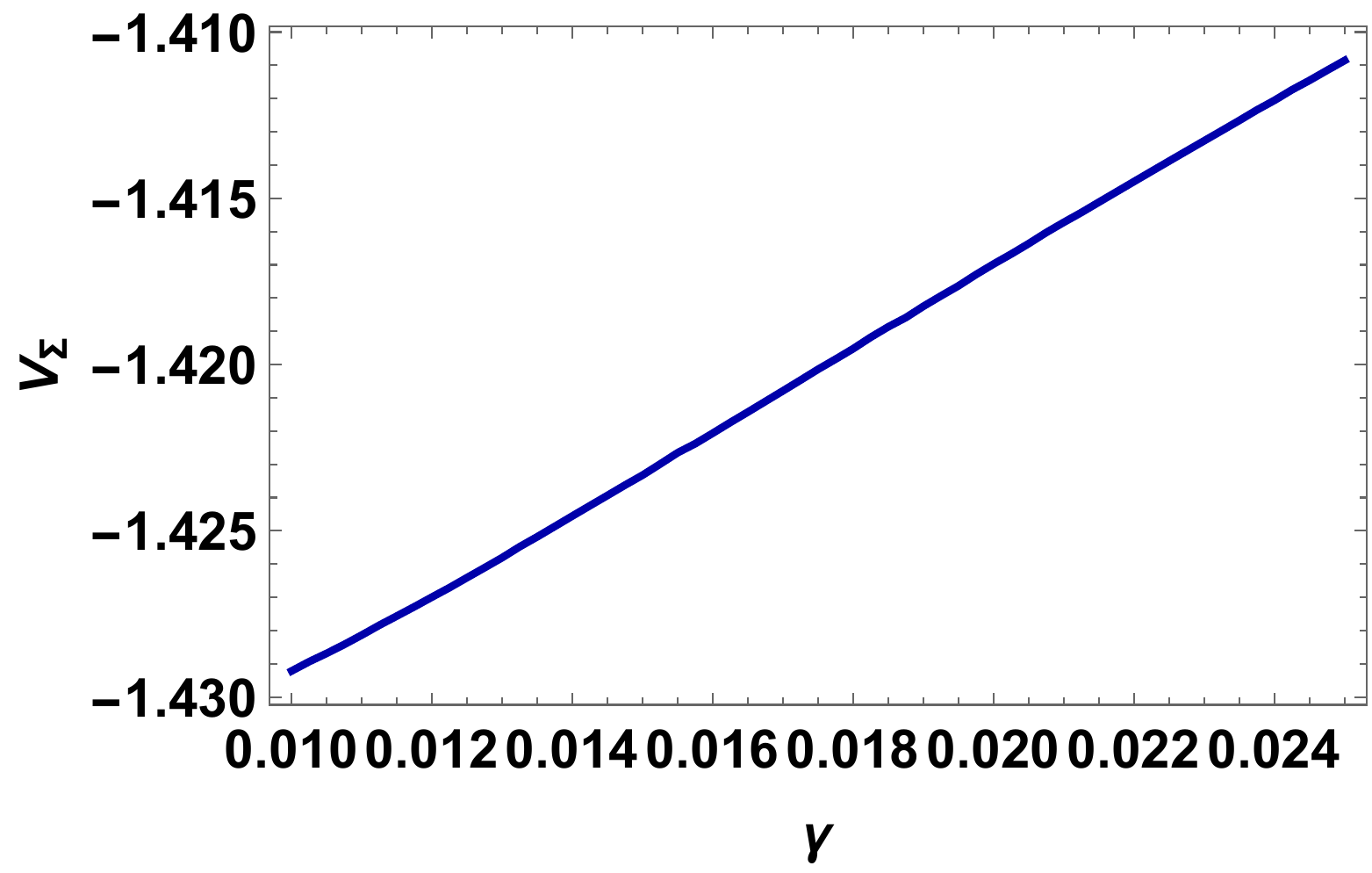}&
\includegraphics[width=0.5\textwidth]{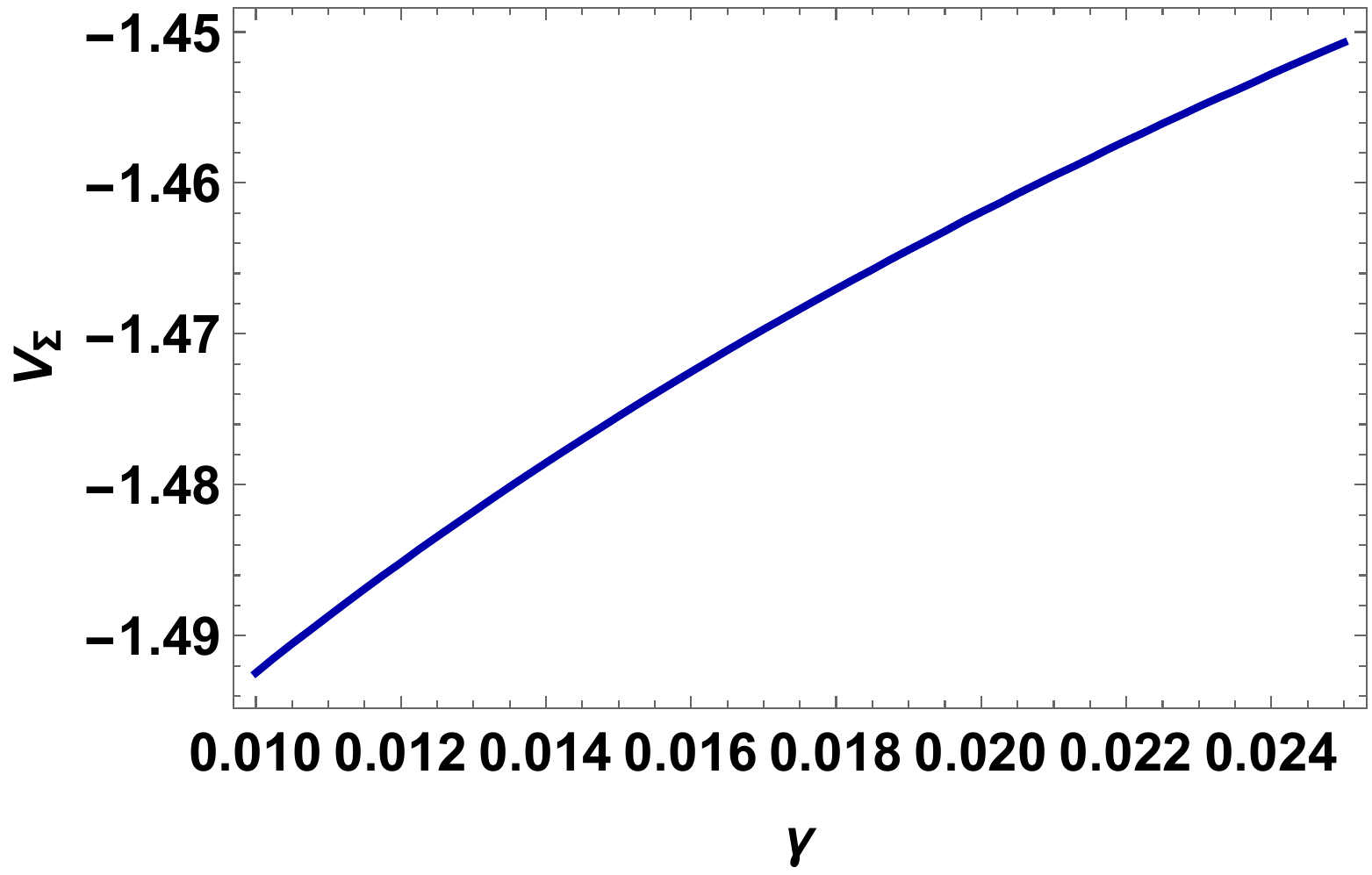}\\
\end{tabular}
\caption{$V_{\Sigma}$ as a function of $\gamma$ for $r_*=-0.04$ (top left), $-0.02$ (top right),
$-0.01$ (bottom left) and $0.01$ (bottom right)} \label{Vvsgamma}}
\end{figure}

We will choose a range of $R$ (with fixed $\gamma/R$) so that 
we get a multi-valued $\ell$ as a function of $r_*$ (see fig.(6) of \cite{MyersRG}). We will focus on $\gamma = 0.01$ for which 
the phase transition has been shown in fig.(\ref{SandV}). In principle, a similar analysis will hold for each value of $\gamma$. 
With this in mind, in fig.(\ref{Vvsgamma}), we have plotted $V_{\Sigma}$ as a function of 
$\gamma$ for $r_*=-0.04$, $-0.02$, $-0.01$, and $0.01$  (top-left, top-right, bottom-left and bottom-right, respectively). 
We have checked
that the behaviour of $V_{\Sigma}$ as a function of $\gamma$ for $r_* < -0.04$ is qualitatively similar to that with $r_*=-0.04$. Similarly,
for $r_* > 0.01$, $V_{\Sigma}$ exhibits the same behaviour as for $r_* = 0.01$.
% % % % % % % % % % % % % % % % % % % % % % % % % % % % % % % % % % % % % % % % % % % % % % % % % 
% % % % % % % % % % % % % % % % % % % % % % % % % % % % % % % % % % % % % % % % % % % % % % % % % 

It can be seen that at $\gamma = 0.01$ (which is our point of interest here), the quantity $dV_{\Sigma}/d\gamma$ changes sign 
as we vary $r_*$ from $-0.02$ to $-0.01$. Since the numerics become somewhat unstable as we 
increase the precision of $r_*$, we could not adopt a finer
resolution. In any case, if we assume that the change of sign in the derivative occurs at $r_* \simeq -0.015$ (somewhere in between
the values of $-0.02$ and $-0.01$), this corresponds to $\ell \approx 1.78$, i.e close to the point $\ell \approx 1.79$ 
at which the phase transition occurs 
(for $\gamma = 0.01$) as seen from fig. \ref{SandV}(a). This is {\it indicative} of the fact that within the limits of numerical error, 
the first derivative $dV_{\Sigma}/d\gamma$ changes sign across the phase transition. 
At this point, we are unable to make a stronger statement regarding the second derivative of $V_{\Sigma}(\gamma)$ which should
be related to the Fisher information. Calculating the numerical derivatives of $V_{\Sigma}(\gamma)$ directly would be
interesting. This certainly deserves further investigation and we will leave this as a future consideration. 

\section{Complexity with the Janus ansatz}

Having computed the complexity for the RG flow in the Poinca\'re slicing, we now proceed to the Janus solution. 
The Janus solution was first introduced in the literature in \cite{Bak} and was a non-supersymmetric deformation of AdS 
resulting in a dual field theory which had a co-dimension  one defect. The defect separated the field theory into two regions 
with different Yang-Mills couplings. Conformal field theories of this type which have an interface or a boundary have been 
extensively studied in the context of holography (see, for example, \cite{Gutperle, Gutperle1, Jensen}). In the subsequent 
discussion, we shall mainly follow the conventions of \cite{Gutperle}. 

The Janus solution is an ansatz where $AdS_{d+1}$ is foliated using $AdS_d$ slices. Let us take the Poinca\'re sliced 
metric of $AdS_{d+1}$,
\be
ds^2 = \frac{1}{z^2} \left( dz^2 + dx_{\perp}^2 - dt^2 + \sum_{i=2}^{d-1} dx_{i}^2 \right)
\ee
The above metric can be mapped to the Janus ansatz by the following transformation,
\be
x_{\perp} = y \cos \mu
\qquad
z = y \sin \mu
\ee
resulting in,
\be
ds^2 = \frac{1}{\sin^2 \mu} \left(d\mu^2 + \frac{dy^2 - dt^2 + \sum_{i=2}^{d-1} dx_{i}^2}{y^2} \right) 
\ee
In contrast to the Poinca\'re sliced metric where the boundary is located at $z=0$, the above metric has three distinct 
components which can be interpreted as boundaries, namely, $\mu=0, \pi$ and $y=0$. We can interpret the $\mu=0, \pi$ as 
half spaces corresponding to two different CFTs and the $y=0$ as the co-dimension one defect where the two half-spaces 
and joined together. Motivated by the prospect of studying how the complexity behaves in a boundary/interface CFT, we 
compute the volume of the Janus solution in the subsequent discussion, restricting ourselves to $d=2$.

Since we are interested in a RG flow scenario, we make the following slight change to the Janus ansatz,
\be
\label{JanusRG}
ds^2 = f(\mu) \left( d\mu^2 + \frac{dy^2 - dt^2 + \sum_{i=2}^{d-1} dx_{i}^2}{y^2} \right) 
\ee
where the factor $f(\mu)$ equals $f(\mu)=\frac{1}{\sin^2 \mu}$ at the fixed points where the space is AdS. To solve for the 
geometry numerically, we shall need the Einstein equations resulting from eq. (\ref{action}), with $\phi \equiv \phi(\mu)$. 
Plugging in the metric (\ref{JanusRG}), we get the following equations \cite{Gutperle},
\bea
\label{Janusequations}
&& \phi'' - f \hat{V}'(\phi) + \frac{d-1}{2}\frac{f'}{f}\phi' = 0 \nn\\
&& \frac{f''}{f} - \frac{3}{2} \frac{f'^2}{f^2} + \frac{4}{d-1}\phi'^2 -2 = 0 \nn\\
&& \frac{1}{4}\phi'^2 - \frac{d(d-1)}{32} \frac{f'^2}{f^2} - \frac{d(d-1)}{8} + \frac{d(d-1)}{8}f - \frac{1}{2}f \hat{V} = 0
\eea
where we have written the potential as,
\be
V(\phi) = -\frac{d(d-1)}{4} +  \hat{V}(\phi)
\ee
Out of the three equations in eqs.(\ref{Janusequations}), we treat the first two as independent equations and the third as a constraint. 
While seeking solutions to the above equations that specify the geometry, we take a toy model for the potential of the form, 
\be
\hat{V}(\phi) = \frac{1}{2} m^2 \phi^2 + \frac{1}{4!}\lambda_{4} \phi^4 
\ee
According to the holographic dictionary, the mass of the scalar field $m$ is related to the conformal dimension in the dual field theory as,
\be
m^2 = \Delta(\Delta-d)
\ee
Since our motivation is to study RG flows, which flow to IR fixed points, we choose the conformal dimensions such that 
the scalar operator is relevant in the IR, i.e,
\bea
-\frac{d^2}{4} < m^2 < 0 \qquad
\frac{d}{2} < \Delta < d
\eea
In the $AdS_d$ slicing, $\phi$ can be shown to have the near boundary behaviour,
\be
\phi(\mu) = \alpha \mu^{\Delta} + \beta \mu^{d-\Delta} + \ldots
\ee
This behaviour is actually analogous to the well-known behaviour of the scalar field the Poinca\'re slicing \cite{Gutperle} and 
$\alpha$ and $\beta$ may be identified with the expectation value and the source of the operator respectively.

With the eqs. (\ref{Janusequations}) fully specified, the initial conditions for $\phi(\mu)$ and $f(\mu)$ can now be specified. 
We take the boundary of one half space to be located at $\mu = 0$. As this is a fixed point in the RG flow, the space in AdS 
and so $f(\mu)=\frac{1}{\mu^2}$ to leading order in $\mu$. Going beyond the leading order, the asymptotic behaviour of the 
fields is given by \cite{Gutperle},
\bea
\label{asympgeom}
&& f(\mu) = \frac{1}{\mu^2} + \frac{1}{3} + \frac{1}{15}\mu^2 -2\beta^2 \frac{\Delta-2}{2\Delta-5} \mu^{2-2\Delta}+\ldots \\
&& \phi(\mu) = \beta \mu^{2-\Delta}-\frac{1}{12}\beta (\Delta-1)\mu^{4-\Delta}+\ldots
\eea
It may be noted here that we have set the expectation value of the scalar field $\alpha=0$.

Similar to the calculation for entanglement entropy, we consider a strip entangling surface of length $\ell$. The minimal surfaces 
are those that intersect the constant time slice of the geometry,
\be
ds_{\Sigma}^2 = f(y) \left(d\mu^2+\frac{dy^2}{y^2} \right)
\ee
and are parametrized by $y \equiv Y(\mu)$. The area functional is given by,
\be
\text{Area} = \int d\mu \sqrt{f(\mu) \left(1+\frac{Y'(\mu)^2}{Y(\mu)^2} \right)} 
\ee
The geodesic equation that minimizes the above functional is given by,
\be
\label{geodesiceqn}
f' Y' (Y^2+Y'^2)+2 f Y (Y Y'' - Y'^2) = 0
\ee 
Using the asymptotic behaviours of $f(\mu)$ and $\phi(\mu)$ from eqs. (\ref{asympgeom}), we get the initial data for $Y(\mu)$ as,
\be
Y(\mu) = \ell+\hat{y}\mu^2 +\frac{\beta^2 \hat{y}(\Delta-2)}{(\Delta-3)(2\Delta-5)}\mu^{6-2\Delta}
\ee 
Now, we proceed to solve the set of eqs. (\ref{Janusequations}) and (\ref{geodesiceqn}) numerically using a shooting method. 
The characteristics of the solutions are discussed at length in \cite{Gutperle}, but we include some brief remarks here for completeness. 
Whether the geometry is dual to an ICFT or a BCFT depends on the source strength $\beta$ of the scalar operator in the conformal 
field theory. It can be found that the solution blows up if the magnitude of $\beta$ is increased beyond a certain critical value (which of 
course, depends on the two other constants which have to be fixed, namely $\lambda_4$ and $\Delta$). Numerically, we monitor the 
value of $f(\mu)$ while integrating the system of equations. For a value of $\beta$ below the critical value, we find that $f(\mu)$ may 
again be approximated by $\frac{1}{\sin^2 \mu}$ at $\mu=\mu_{*}$, indicating that the geometry has flowed into another AdS region. 
This is thus an example of an ICFT. For values of $\beta$ greater than the critical value, $f(\mu)$ blows up at $\mu=\mu_{*}$, indicating 
that the theory has acquired a mass. 

The geodesic solutions are also obtained from eqs.(\ref{Janusequations}) and (\ref{geodesiceqn}) and it may be readily verified that 
for the BCFT geometry, which is singular at $\mu=\mu_{*}$, there is only one geodesic, $Y(\mu)=\ell$ which reaches the singularity. All 
other geodesics are repelled and do not reach the singularity. So, this unique geodesic is the one which is used to calculate the complexity. 
In the ICFT case of course, there is an infinite class of geodesics which reach the point $\mu=\mu_{*}$.

With the numerical solutions of $f(\mu)$ and $Y(\mu)$ in hand, we proceed to calculate the complexity, which is given by
\be
V_{\Sigma} = \int_{\mu_{\epsilon}}^{\mu_{*}} d\mu f(\mu) \int_{\epsilon}^{Y(\mu)} \frac{dy}{y} 
\label{JanusVolume}
\ee
Here, $\mu_{\epsilon}=\frac{\epsilon}{\ell}$ is the counterpart of the UV cut-off in the Poinca\'re slicing. Following \cite{Miller},\cite{Bak1},
we note that strictly speaking, we should impose a separate cut-off (say $\delta$), for the coordinate $y$. However, since we concentrate
on a purely numerical treatment (and moreover we have set $\ell = 1$ throughout) it is not difficult to see that this does not affect any
of our results below. 

Similar to the entanglement entropy, 
the complexity is a divergent quantity due to the the behaviour $f(\mu) \sim \frac{1}{\mu^2}$ as $\mu \rightarrow 0$. The expression for the 
volume may be regulated by subtracting out the divergent part, which may be evaluated to be $\frac{\ell}{\epsilon} \log(\frac{\ell}{\epsilon})$. 
By inspection of eq.(\ref{JanusVolume}), it is seen that the finite part of the complexity is independent of $\ell$. 
% % % % % % % % % % % % % % % % % % % % % % % % % % % % % % % % % % % % % % % % % % % % 
\begin{figure}[htp]
\centering
\includegraphics[width=.3\textwidth]{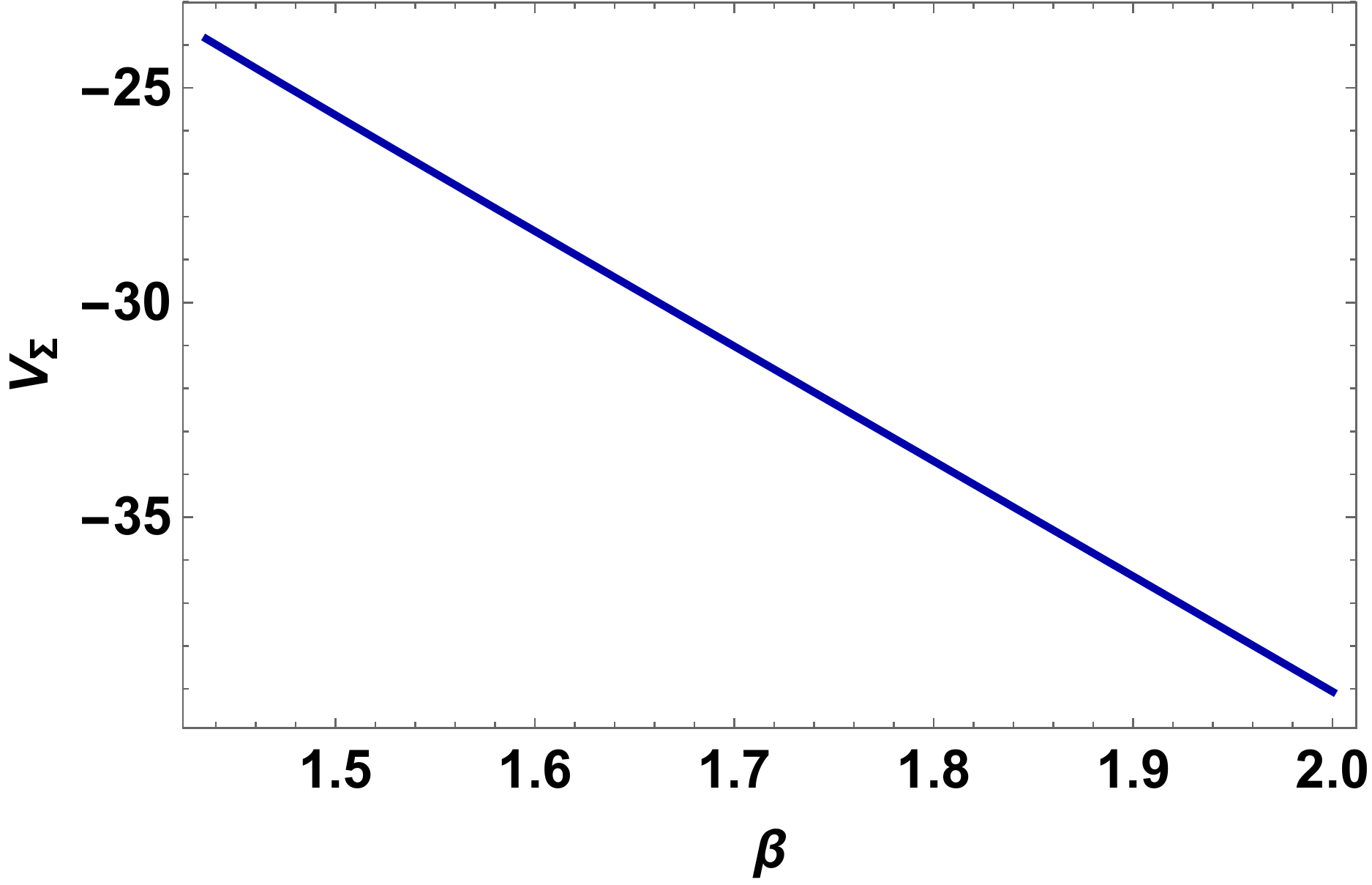}\hfill
\includegraphics[width=.3\textwidth]{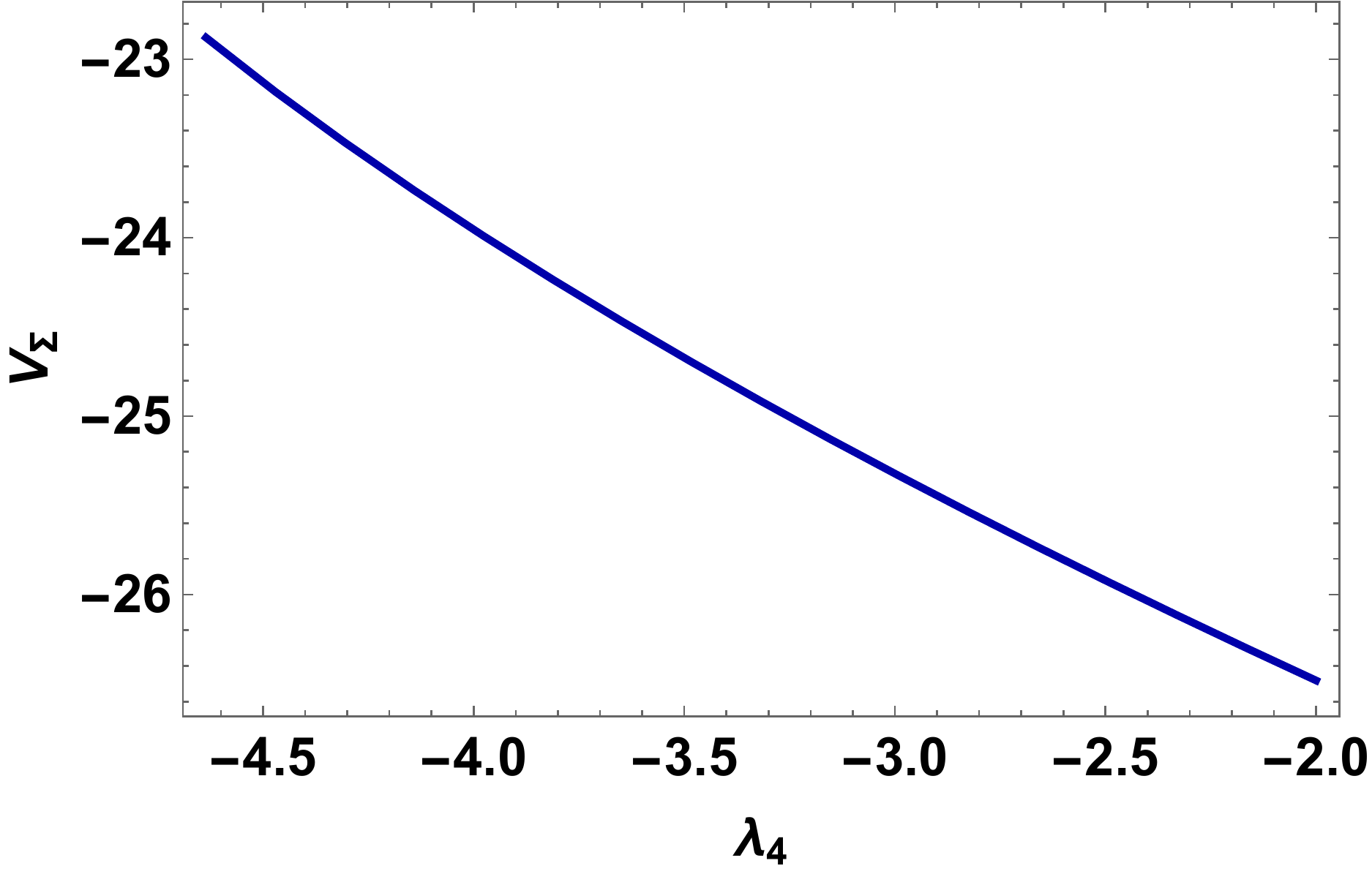}\hfill
\includegraphics[width=.3\textwidth]{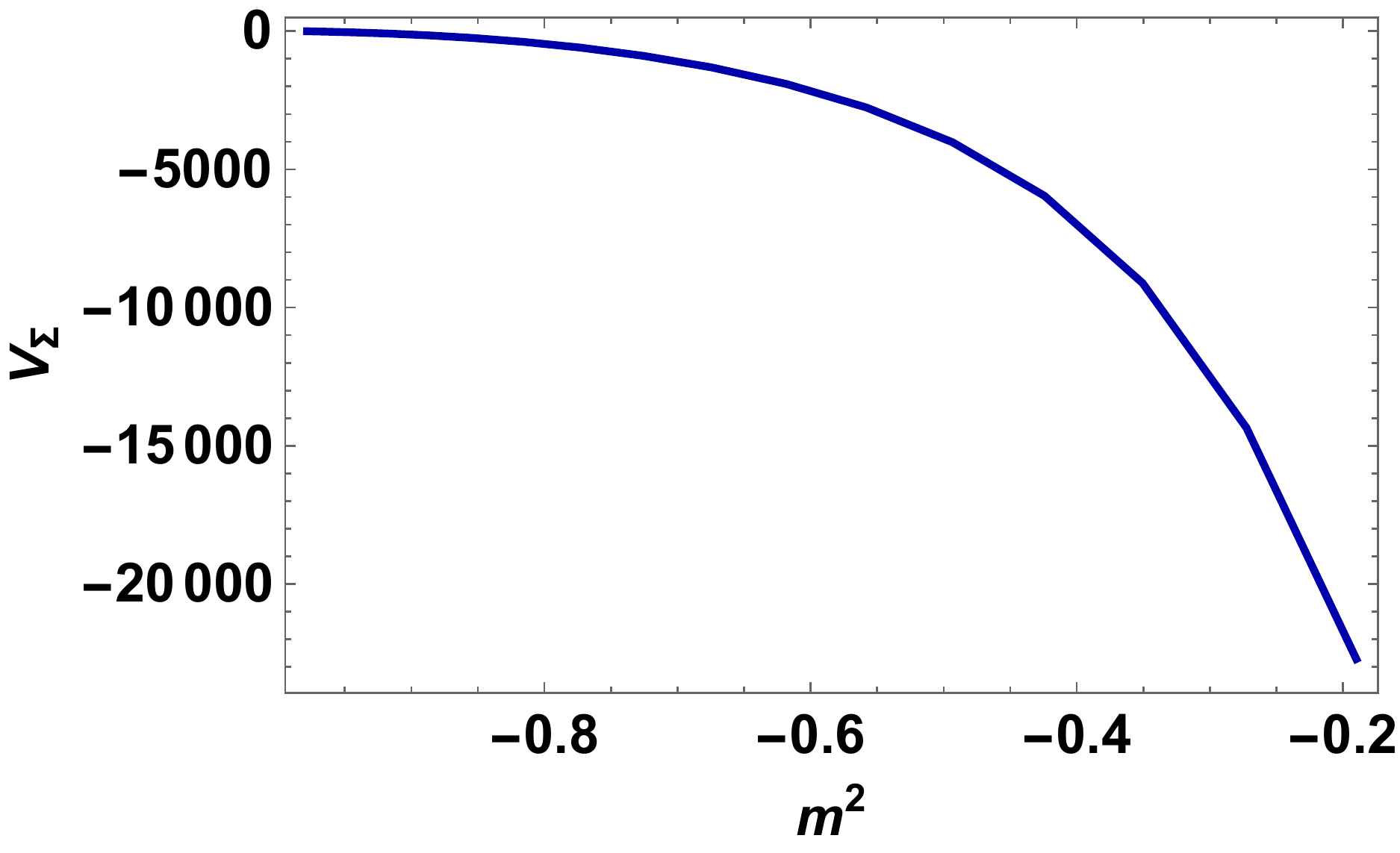}
\caption{$V_{\Sigma}$ as a function of (a) $\beta$, (b) $\lambda_{4}$ and (c) $m^2$ for BCFTs. For (b) and (c), $\beta=1.4$.}
\label{BCFTgraphs}
\end{figure}
% % % % % % % % % % % % % % % % % % % % % % % % % % % % % % % % % % % % % % % % % % % % % % % % 
% % % % % % % % % % % % % % % % % % % % % % % % % % % % % % % % % % % % % % % % % % % % % % % % 
\begin{figure}[htp]
\centering
\includegraphics[width=.3\textwidth]{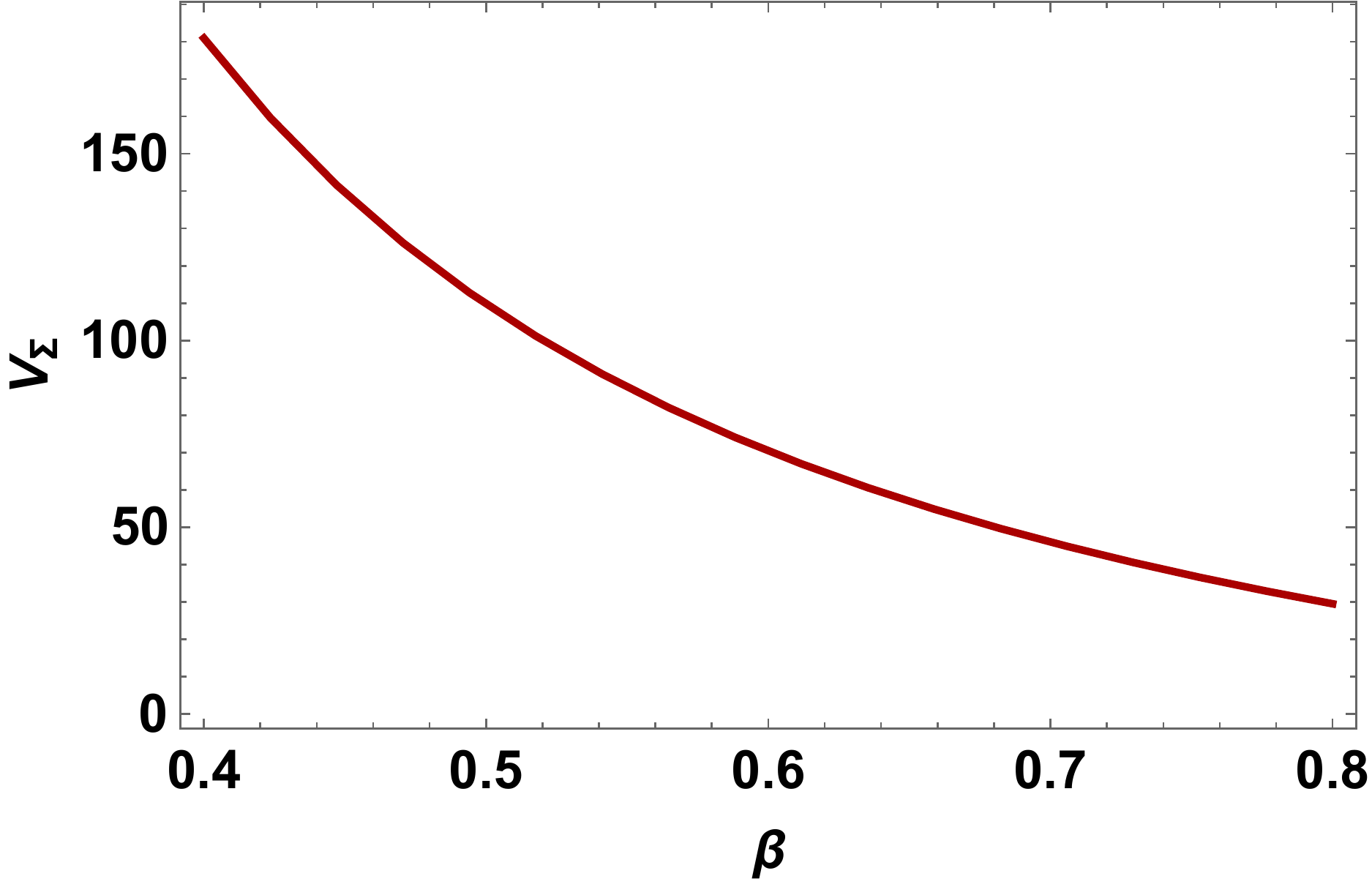}\hfill
\includegraphics[width=.3\textwidth]{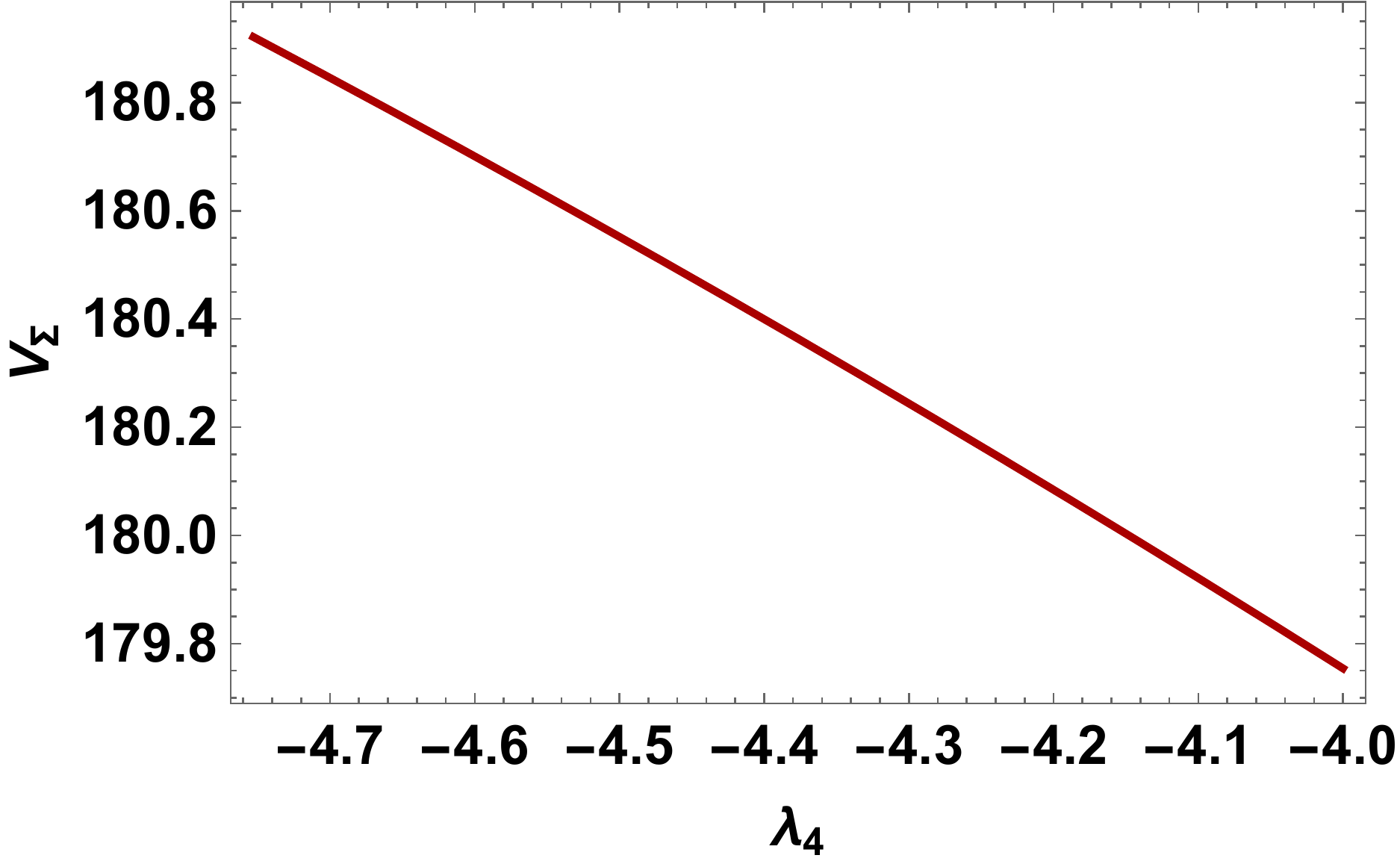}\hfill
\includegraphics[width=.3\textwidth]{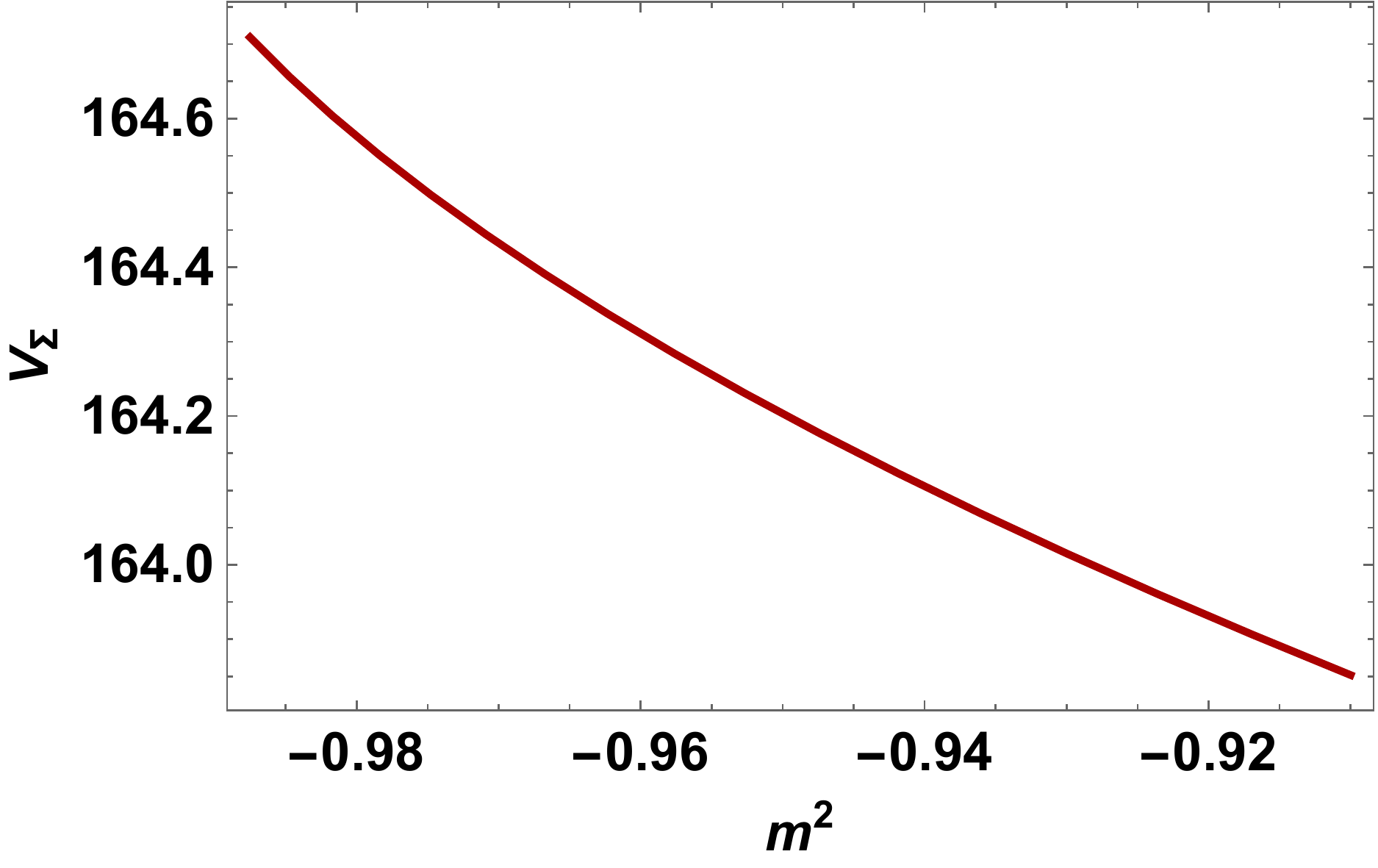}
\caption{$V_{\Sigma}$ as a function of (a) $\beta$, (b) $\lambda_{4}$ and (c) $m^2$ for ICFTs. For (b) and (c), $\beta=0.4$.}
\label{ICFTgraphs}
\end{figure}
% % % % % % % % % % % % % % % % % % % % % % % % % % % % % % % % % % % % % % % % % % % % % % % % 
  
In figs. (\ref{BCFTgraphs}) and (\ref{ICFTgraphs}), we have plotted $V_{\Sigma}$ as a function of $\beta$, $\lambda_4$ and $m^2$. Following
our discussion in section 2, The latter two quantities can indeed be thought of as coordinates on a parameter manifold. Hence, these figures
can be viewed as pictorial depictions of analogues of the Fisher metric of section 2. However, we should point out that 
numerical limitations did not allow us to take small values of the couplings that we would ideally have liked to. Note also that the 
complexity is a monotonically decreasing function of $\beta$. This is similar in nature to the behaviour of the entanglement entropy
(or more appropriately the ``g function'') studied for BCFTs in $d=2$ \cite{Gutperle}.
    
\section{Conclusions and discussions}

In this paper, we have studied subregion holographic complexity for renormalization group flow scenarios. We have considered two
distinct cases here : first, we studied this quantity for a domain wall setup, both with a sharp and a smooth domain wall. Next, we computed
the complexity for the case of the Janus solution. In addition to the inherent importance of calculating subregion complexity as it might result in
newer insight into the gauge/gravity duality, appropriate derivatives of the complexity can be thought of 
as defining the Fisher information metric for these scenarios in lines with \cite{Souvik}. 

Our computation of the complexity or the RT volume for a sharp domain wall scenario revealed some interesting properties, 
that could be contrasted with the behaviour of the 
entanglement entropy. This analysis was done both for disc shaped and strip shaped
entangling regions. In particular, we noticed that the holographic phase transition revealed by the analysis of \cite{MyersRG} cannot be 
captured by the complexity. However, the
situation changed when we computed the RT volume for the smooth domain wall scenario, and we found that interestingly, the complexity 
captured the physics of phase transition here, and our results are indicative of the fact that the derivative of the complexity (related to the 
Fisher metric) changes sign at the phase transition. Finally, we computed the complexity for the Janus solution and obtained its variation with
respect to the system parameters, which can in principle be used to determine the information metric. 

A deeper understanding of subregion complexity and its relation to the Fisher information metric is of great interest. We have taken the initial steps
in this paper, and hope to report on further progress in a forthcoming work. 

\bibliography{references.bib}
\bibliographystyle{JHEP}
\end{document}